\documentclass[aps,prc,reprint,showpacs,groupedaddress,onecolumn]{revtex4-1}

\usepackage{graphicx,color} 
\usepackage{dcolumn}  
\usepackage{bm}       
\usepackage{amsmath}
\usepackage{scalerel,stackengine}

\begin{document}

\newcommand {\nc} {\newcommand}

\newcommand{\vv}[1]{{$\bf {#1}$}}
\newcommand{\ul}[1]{\underline{#1}}
\newcommand{\vvm}[1]{{\bf {#1}}}
\def\btau{\mbox{\boldmath$\tau$}}

\nc {\IR} [1]{\textcolor{red}{#1}}
\nc {\IB} [1]{\textcolor{blue}{#1}}
\nc {\IP} [1]{\textcolor{magenta}{#1}}
\stackMath
\newcommand\reallywidehat[1]{%
\savestack{\tmpbox}{\stretchto{%
  \scaleto{%
    \scalerel*[\widthof{\ensuremath{#1}}]{\kern-.6pt\bigwedge\kern-.6pt}%
    {\rule[-\textheight/2]{1ex}{\textheight}}
  }{\textheight}%
}{0.5ex}}%
\stackon[1pt]{#1}{\tmpbox}%
}

\title{$^6$Li in a Three-Body Model with Realistic Forces: Separable vs.
Non-separable Approach}

\author{L.~Hlophe$^{(a)}$}
\author{Jin Lei$^{(b)}$}
\author{Ch.~Elster$^{(b)}$}
\author{A. Nogga$^{(c)}$}
\author{F.M.~Nunes$^{(a)}$}

\affiliation{(a) National Superconducting Cyclotron Laboratory and Department of Physics and Astronomy, Michigan State University, East Lansing, MI 48824, USA \\
(b)Institute of Nuclear and Particle Physics,  and
Department of Physics and Astronomy,  Ohio University, Athens, OH 45701,
USA \\
(c) IAS-4, IKP-3, JHCP, and JARA-HPC,  Forschungszentrum J\"ulich, D-52428
J\"ulich, GER
}

\date{\today}

\begin{abstract}
\begin{description}
\item[Background] 
Deuteron induced reactions are widely used to probe nuclear structure
and astrophysical information.
Those (d,p) reactions may be viewed as three-body reactions and described with Faddeev techniques.

\item[Purpose] Faddeev equations in momentum space have a long tradition of
utilizing separable interactions in order to arrive at sets of coupled integral
equations in one variable. However, it needs to be demonstrated that 
their solution based on separable interactions  agrees exactly with
solutions based on non-separable forces. 

\item[Results] 
The ground state of $^6$Li is calculated via momentum space Faddeev
equations using the
CD-Bonn neutron-proton force and a Woods-Saxon type neutron(proton)-$^4$He
force. For the latter the Pauli-forbidden $S$-wave bound state is 
projected out. This result is compared to a calculation in which the 
interactions in the two-body subsystems are represented by 
separable interactions  derived in the Ernst-Shakin-Thaler framework.   

\item[Conclusions] We find that calculations based on the separable
representation of the interactions and the original interactions give
results that agree to four significant figures for the binding energy, provided an
off-shell extension of the EST representation is employed in both
subsystems. The momentum distributions computed in both approaches also fully agree with each other.
\end{description}
\end{abstract}


\maketitle

\section{Introduction}
\label{intro}

A variety of applications of nuclear physics require the understanding of neutron
capture on unstable nuclei. Due to the short lifetimes involved, direct measurements
are currently not possible, and thus indirect methods using (d,p) reactions have been
used for both the direct capture \cite{jolie, Kozub:2012ka} as well as the compound \cite{RevModPhys.84.353} components.
A recent review on (d,p) reactions and its connection to neutron capture can be found in \cite{Potel:2017z}. 
In addition, single neutron transfer (d,p) reactions can be used to constrain proton capture cross sections, due to mirror symmetry (e.g. \cite{Pain:2015}).
Beyond these astrophysical motivations, single-nucleon transfer reactions involving the deuteron
have  been the preferred tool to study shell evolution in nuclear structure, both for nuclei close and far from stability (see
Refs.~\cite{Schmitt:2012bt,Jones:2011kp} for two recent examples).
In all these cases, a reliable reaction theory for (d,p) is a critical ingredient.

Scattering and reaction processes involving deuterons either as projectile or
as target are perhaps the most natural three-body problem in the realm of nuclear reactions. 
The binding energy of the deuteron is so small that its root-mean-square radius is significantly 
larger than the range of the force. That means that when a deuteron interacts with a compact, well bound
nucleus, one may expect that it will behave like a three-body system consisting of proton $p$, 
a neutron $n$, and a nucleus $A$. The obvious three-body reactions are elastic scattering, 
rearrangement and breakup processes. 
In order to describe those processes on the same footing, deuteron-nucleus
scattering should be treated at least at the three-body level. Note that if the target itself has low-lying excitations, one may need to go beyond the pure three-body treatment, see  e.g. \cite{Deltuva2013}.
However, for the application we consider here (namely $^6$Li$~\equiv n+p+\alpha$), one expects the three-body treatment to be sufficient. 

The three-body Hamiltonian governing the dynamics of the (d,p) reactions contains the well understood nucleon-nucleon
(NN) interaction as well as an effective interaction between the nucleons and the target.
Commonly these nucleons-nucleus interactions are parameterized by  phenomenological 
optical potentials which fit a large body of elastic scattering data~\cite{Varner:1991zz,Weppner:2009qy,Koning:2003zz}.

The application of momentum space Faddeev techniques to nuclear reactions has been pioneered in Ref.~\cite{Deltuva:2009fp}, and successfully applied to (d,p) reactions for light nuclei~\cite{Deltuva:2009cr}. However, when extending these
calculations to heavier nuclei~\cite{Nunes:2011cv,Upadhyay:2011ta}, it becomes
 apparent that the screening techniques employed for incorporating the Coulomb interaction in Faddeev-type
reaction calculations with light nuclei cannot be readily extended to the heaviest nuclei. Therefore, a new method for treating 
(d,p) reactions with the exact inclusion of the Coulomb force
 as well as target excitation was formulated in Ref.~\cite{Mukhamedzhanov:2012qv}. 
This new approach relies on a separable representation of the pairwise forces. 

Separable representations of the forces between constituents forming the subsystems
in a Faddeev approach have a long tradition in few-body physics.
In the context of describing light nuclei like  
$^6$Li~\cite{Eskandarian:1992zz,Lehman:1982zz,Lehman:1982zz,Lehman:1978zz} and 
$^6$He~\cite{Ghovanlou:1974zza} in a three-body approach, rank-1 separable interactions of Yukawa-type
 have been successfully used. In the case of the three-nucleon problem,  separable representations for the NN force
of higher rank had to be developed in order  to improve the agreement with exact Faddeev calculations~\cite{Haidenbauer:1982if,Haidenbauer:1986zza}. Those
were based on the scheme suggested by Ernst-Shakin-Thaler~\cite{Ernst:1973zzb} (EST). 

The pioneering work of Hlophe and collaborators
\cite{Hlophe:2013xca,Hlophe:2014xca,Hlophe:2015rqn} demonstrated that an
EST-based separable interaction of rank up to 5 provides a precise description
of nucleon-nucleus elastic scattering for a wide range of energies, including
nuclei as heavy as $^{208}$Pb. The development of these separable complex (and
energy dependent) effective  potentials opens the path to apply the method of Ref.~\cite{Mukhamedzhanov:2012qv} to the  three-body  $A(d,p)B$ reaction problem.

Since a separable expansion of the nuclear transition amplitudes can be viewed as a basis expansion, 
it is critical to understand the convergence of the numerical results. In order to benchmark these calculations, one needs to compare to the solution of the problem without the use of separable interactions. Such a comparison was successfully carried out for
neutron-deuteron scattering at 10~MeV \cite{Cornelius:1990zz}, and at slightly higher energies in
Ref.~\cite{Nemoto:1998wt}. Both studies showed that, for a converged expansion of the force in
the two-body subsystems, observables in the three-body system agree. 

For the applications we have in mind, the benchmarks need to be performed for
the $A+d$ case in a regime for which non-separable solutions are possible and
exact.  Furthermore, our work aims to establish that the approach based on 
separable two-body transition matrices  is {\it equivalent} to the approach using those transition matrices directly, given the convergence in the expansion. 
We choose as benchmark, the ground state
of $^6$Li because there is a large number of reference calculations in the
literature; our goal is for an agreement between the separable and the
non-separable approach of up to four significant figures in the binding
energy. The ultimate goal is to apply the separable approach to nuclear
reactions. Here we expect to lose some precision in solving the Faddeev equations in the continuum. Note that benchmarks performed for the four-nucleon bound state ensured 4 digit accuracy \cite{benchmark1} while the corresponding work for positive energies provided only a 2-digit accuracy \cite{benchmark2}.
This should also be sufficient for the problem we are considering, particularly when computing (d,p) observables.

In this work, our EST-based separable expansion uses off-shell transition amplitudes at negative energies as basis states. Those states depend on two
parameters, namely the energy and the off-shell asymptotic momentum, which are chosen independently. This is in contrast to previous work on the neutron-deuteron system ~\cite{Haidenbauer:1982if,Haidenbauer:1986zza}, which did not explore the full parameter space. 
 The effective interactions in the neutron-alpha and proton-alpha channels 
are given by a Woods-Saxon type potential fitted to phase shifts in the S- and
P-wave channels.
Since 
the $n\alpha$ Woods-Saxon potential supports a bound state in the S-wave two-body channel, which is 
Pauli forbidden, we derive a projection scheme for both approaches
which differs from previous works \cite{Lehman:1982zz,Schellingerhout:1993ku} in that
it does not modify the underlying Faddeev equations.

In Sec. II a brief summary of the theory is provided, including the three-body equations we solve and the new formulation used to
project out the Pauli forbidden S-wave state in the neutron(proton)-alpha channel.
The inputs to the problem are presented in Sec.~\ref{subsystems}, including the interactions that govern the  two-body subsystems,
and the results for the $^6$Li binding energy and wavefunction are discussed in
Sec.~\ref{separable} and~\ref{properties}. 
 Our findings are summarized in Sec.~\ref{summary}.


\section{Formal Considerations}
\label{formal}

\subsection{Faddeev Equations for the ground state of $^6$Li}
\label{faddeevequations}

The bound state of three particles with masses $m_i$, $m_j$, and $m_k$ and spins $j_i$, $j_j$,
and $j_k$ which interact via pairwise forces $V^i \equiv V_{jk}$  ($i,j,k=1,2,3$ and cyclic
permutations thereof) is given by the Schr\"odinger equation, which reads in integral form
\begin{equation}
|\Psi\rangle = G_{0}(E_3) \sum^{3}_{i=1}V^{i} |\Psi\rangle.
\label{eq:1.1}
\end{equation}
Here the free propagator is given by
 $G_{0}(E)=(E_3-H_{0})^{-1}$, where $H_0$ stands for the free Hamiltonian
and $E_3$ for the binding energy of the three-body system. Introducing Faddeev components
\begin{equation}
|\Psi\rangle = \sum_{i=1}^3 |\psi_{i}\rangle \equiv |\psi_{jk}\rangle + |\psi_{ki}\rangle
+|\psi_{ij}\rangle, 
\label{eq:1.2}
\end{equation}
with
\begin{equation}
|\psi_{i}\rangle =  G_{0}(E_3) V^{i} |\Psi\rangle,
\label{eq:1.3}
\end{equation}
leads to three coupled integral equations for the three components $|\psi_{i,j,k}\rangle$, 
\begin{eqnarray}
|\psi_{i}\rangle = G_{0}(E_3)\;t_{i}\sum_{j\neq i} |\psi_{j}\rangle.
\label{eq:1.4}
\end{eqnarray}
The operator $t_i = V^i + V^i G_0(E_3) t_i$ describes the two-body t-matrix in the subsystem $jk$. 
In order to solve Eqs.~(\ref{eq:1.4}) standard Jacobi momenta are used,
\begin{eqnarray}
\vec{p}_k &\equiv& \vec{p}_{ij} = \mu_{ij} \left(
       \frac{\vec{k}_i}{m_i}-\frac{\vec{k}_j}{m_j}\right) \cr
\vec{q}_k &\equiv& \vec{q}_{ij} = \mu_{3b,k} \left(\frac{\vec{k}_k}{m_k} -
\frac{\vec{k}_i+\vec{k}_j}{m_i +m_j} \right).
\label{eq:1.5}
\end{eqnarray}
Here the two-body reduced mass $\mu_{ij}$ and the three-body reduced mass $\mu_{3b,k}$ are
given by
\begin{eqnarray}
\mu_{ij}&=&\frac{m_{i}m_{j}}{m_{i}+m_{j}} \cr
\mu_{3b,k} &=&\frac{m_{k}(m_{i}+m_{j})}{M} ,
\label{eq:1.6}
\end{eqnarray}
with $M=m_{i}+m_{j}+m_{k}$ being the total mass of the system. 

Instead of using a three-dimensional Jacobi basis, we expand into momentum eigenstates which
depend on the magnitude of the momenta and angular momentum eigenstates. The orbital angular
momenta of the three particles are coupled to total angular momentum $J$ and its third component $M_J$,
\begin{equation}
|{p}_k{q}_k \alpha_{k} \rangle_{(ij)k} = |p_kq_k((l_k(j_ij_j)s)J_{ij} (\lambda_k j_k)J_k) JM_J
\rangle_{(ij)k},
\label{eq:1.7}
\end{equation}
which are normalized as
\begin{equation}
{}_{(ij)k}\langle {p}'_k{q}'_k \alpha_{k} ' | {p}_k{q}_k \alpha_{k} \rangle_{(ij)k}=
\frac{\delta(p'_k-p_k)}{p'_kp_k} \frac{\delta(q'_k-q_k)}{q'_kq_k} \delta_{\alpha_{k} '\alpha_{k}}.
\label{eq:1.8}
\end{equation}
The notation $(ij)k$ indicates that $k$ is the spectator.

Since we represent each Faddeev component $|\psi_{k}\rangle$ in its natural set of Jacobi coordinates $|{p}_k{q}_k \alpha_{k} \rangle_{(ij)k}$,
a transformation between the sets $(jk)i$ to $(ij)k$ and $(ki)j$ to
$(ij)k$ is required. The partial wave representation of these transformations can be calculated 
as outlined in \cite{Balian:1969sd} and can be written as an integral over the cosine x of the relative angle of $p_{k}$ and $q_{k}$
of the Faddeev components evaluated at shifted momenta $\pi'_j = \pi'_j(p'_{k}q_k x) $ and $\chi'_j = \chi'_j (p'_{k}q_k x) $. 
All geometrical information can be parameterized by functions ${\cal G}_{\alpha'_k \alpha'_j}(p'_{k}q_k x)$. 
We give more details on these transformations in Appendix~\ref{appendixA}.

Inserting complete sets of states in Eqs.~(\ref{eq:1.4}) and making use of the geometrical coefficients ${\cal G}_{\alpha'_k \alpha'_j}(p'_{k}q_k x)$, we arrive at a set of three coupled Faddeev equations:
\begin{eqnarray}
\psi_{k}^{\alpha_k}(p_k, q_k) &=&  G_0(E_{q_k};p_k) 
   \sum_{\alpha_k'} \int dp'_k p'^2_k\; t^{\alpha_k \alpha'_k}_{k} (p_k, p'_{k};E_{q_k})  \cr 
   & & \quad \times \int_{-1}^{1} dx \,  \Bigg[
  \sum_{\alpha_i'} {\cal G}_{\alpha'_k \alpha'_i}(p'_{k}q_k x) \;\psi_{i}^{\alpha'_i} (\pi'_i,\chi'_i)+ \sum_{\alpha_j'} {\cal G}_{\alpha'_k \alpha'_j}(p'_{k}q_k x)\;\psi_{j}^{\alpha'_j} (\pi'_j, \chi'_j)\Bigg],
\label{eq:1.9}
\end{eqnarray}
where we introduced the pair kinetic energy $E_{q_k} = E_3-\frac{q_k^2}{2\mu_{3b,k}}$ and the free three-body propagator 
\begin{equation}
G_0(E_{q_k};p_k)  = \frac{1}{E_{q_k}-\frac{p_k^2}{2\mu_{ij}}} \ .
\end{equation}
The two-body t-matrix $t^{\alpha_k \alpha'_k}_{k}$ in the Jacobi coordinates $(ij)k$
is given by the Lippmann-Schwinger equation (LSE), 
\begin{equation}
t_{k}^{\alpha_{k} \alpha_{k}'} (p_k, p'_k;  E_{q_k}) = V^{k; \alpha_{k} \alpha_{k}'} (p_k, p'_k) 
 + \sum_{\alpha''} \int \mathrm{d} p''_k p''^2_k \;
V^{k;\alpha_{k} \alpha_{k}''} (p_k, p''_k) \ G_0(E_{q_k};p_k'') \ 
t_{k}^{\alpha_{k}''\alpha_{k}'} (p''_k, p'_k; E_{q_k}).
\label{eq:1.10}
\end{equation}
For brevity, we labeled the partial wave channels using three-body quantum numbers $\alpha_{k}$. 
Since the LSE corresponds to a two-body problem at an off-shell energy $E_{q_k}$, the interactions 
and $t$~matrices will only dependent on quantum numbers of the two-body subsystems and 
will be diagonal in the spectator quantum numbers. 

As is well known (see e.g. \cite{SchmidZiegelmann}), if the $t$~matrix in the subsystems is
separable, 
\begin{equation}
t_{k}^{\alpha_{k} \alpha_{k}'} (p_k, p'_k;  E_{q_k}) = \sum_{mn} h^{\alpha_{k}}_m(p_k)\; \tau_{mn}^{\alpha_{k} \alpha_{k}'}(E_{q_k})\;
h^{\alpha_{k}'}_n(p'_k),
\label{eq:1.11}
\end{equation}
 the coupled integral equations in two variables, Eqs.~(\ref{eq:1.9}), can be reduced to coupled integral equations in one variable \big(see Eq.~(A5)\big). The indices $\{m,n\}$ represent the rank of the separable potential, and $k$ stands for the index of the Faddeev component.
It is possible, with an appropriate choice of
integration variables and the introduction of the modified  geometric functions ${\cal \tilde
G}_{\alpha'_k \alpha'_j}(q_iq_k x)$ as defined by Eq.~(A6), to obtain a
  separable form for the Faddeev amplitudes:
\begin{equation}
\psi^{\alpha_k}_{k}(p_k ,q_k) \equiv  G_{0}(E_{q_k},p_k)\;\sum\limits_{m} 
 h_m^{\alpha_k}(p_k)\;  F_{m\alpha_k}^{(k)}(q_k).
\label{eq:1.12}
\end{equation}
Reinserting the above expressing into Eqs.~(\ref{eq:1.9}) leads to a coupled set of equations
for the amplitudes  $F_{m\alpha_k}^{(k)}(q_k)$,
\begin{eqnarray}
 F_{m\alpha_k}^{(k)}(q_k)&=& 
\sum\limits_{\nu\alpha_i'}\;\int d\tilde q_i\;\tilde q_i^2\;
\Bigg[\sum\limits_{n\alpha_k'}\;\tau^{\alpha_{k}\alpha_{k}'}_{mn}(E_{q_k})\;Z^{(ki)}_{n\alpha_k',\nu\alpha_i'}(q_k,\tilde q_i)\Bigg]\; 
 F_{\nu\alpha_i'}^{(i)}(\tilde q_i)\cr\cr\cr
  &+&\sum\limits_{\nu\alpha_j'}\;\int d\tilde q_j\;\tilde q_j^2\;
  \Bigg[\sum\limits_{n\alpha_k'}\;\tau^{\alpha_{k}\alpha_{k}'}_{mn}(E_{q_k})\;Z^{(kj)}_{n\alpha_k',\nu\alpha_j'}(q_k,\tilde q_j)\Bigg]\;
F_{\nu\alpha_j'}^{(j)}(\tilde q_j),
\label{eq:1.13}
\end{eqnarray}
where all amplitudes are generated by cyclic permutations of ($ijk$).
The functions
$Z^{(ki)}_{n\alpha_k',\nu\alpha_k'}(q_k,\tilde q_i)$ are the so-called transition
amplitudes~\cite{SchmidZiegelmann} coupling the different types of subsystems. For completeness,
the expressions are explicitly given in Appendix~\ref{appendixA}. We solve both sets of Faddeev
equations using iterative Lanczos-type techniques~\cite{Saad:2003}.

\subsection{Treatment of Pauli Blocking in the Faddeev Equations}
\label{pauliblocking}

Three-body models of nuclei or nuclear reactions require taking Pauli blocking into account to
remove components of the wavefunction that would disappear under full anti-symmetrization of the
(A+2)-body problem. Though this topic has already been extensively treated in the
literature (see
e.g.~\cite{Kukulin:1978he,Thompson:2000ny,Schellingerhout:1993ku,Bang:1983xpz,Lehman:1982zz}),
we need to pick it up again and develop a formulation for projecting out a Pauli forbidden state
in momentum space Faddeev equations that works for separable and non-separable forces alike.

Let us assume that the Pauli forbidden state is created
by a potential $V^i$ in the subsystem $i$. This two-body 
bound state with the wave function
$|\phi^i\rangle$ is a normalized
eigenstate of $H^i=H_0+V^i$. 
It can be projected out by introducing
the channel Hamiltonian
\begin{equation}
\tilde{H}^i = H_0 + V^i + \hat{V}^i  = H^i +\tilde{V}^i, 
\label{eq:1.14}
\end{equation}   
where $\hat{V}^i = \lambda |\phi^i\rangle \langle \phi^i|$ with $\lambda$ being  a large number. 
The Faddeev equations require two-body transition matrices as
input. Thus one needs
\begin{equation}
\tilde{t}_i(z) = \tilde{V}^i + \tilde{V}^i G_0(z) \tilde{t}_i(z), 
\label{eq:1.15}
\end{equation}
with $G_0(z)$ being the free resolvent with $z=E+i\varepsilon$. In this derivation we will drop the subscript $i$, representing the arrangement
channel, for brevity. The discussion is general for each pair that contains forbidden states.
Using the Gell-Mann-Goldberger
relation~\cite{GellMann:1953zz} in the form
\begin{equation}
\tilde{t}(z) =t(z) + \big(1+VG(z)\big) \; \hat{t}(z) \; \big(1+G(z) V\big),
\label{eq:1.16}
\end{equation} 
where $G^{-1}(z) =(z-H)$, and $\hat{t}(z)$ an operator fulfilling the LSE,
\begin{equation}
\hat{t}(z) = \hat{V} + \hat{V} G(z) \hat{t}(z)
 = \hat{V} +\hat{V} \tilde{G}(z)\hat{V}.
\label{eq:1.17}
\end{equation}
Here $\tilde{G}^{-1}(z) = (z-\tilde{H})$.
Since $\hat{V}$ is
separable and of rank-1, the analytic solution for $\hat{t}$ is separable and of rank-1,
\begin{equation}
\hat{t}(z) = |\phi\rangle \;\frac{1}{\frac{1}{\lambda} -\langle \phi|G(z)|\phi \rangle}\; \langle\phi|.
\label{eq:1.18}
\end{equation}
Using $V G(z) = t(z) G_0(z)$ Eq.~(\ref{eq:1.16}) becomes
\begin{equation}
\tilde{t}(z) =t(z) + \frac{|\eta (z)\rangle \langle \bar{\eta} (z)|}
                {\frac{1}{\lambda} - \frac{1}{z-E_b}},
\label{eq:1.19}
\end{equation}
where $E_b$ represents the two-body energy for the bound state $b$ that needs to be projected out, and
\begin{eqnarray}
|\eta (z) \rangle &=& \big(1+t(z)G_0(z)\big)|\phi\rangle \; , \cr
\langle \bar{\eta}(z) | &=& \langle \phi| \big(1+G_0(z)t(z)\big).
\label{eq:1.20}
\end{eqnarray}
Equation~(\ref{eq:1.19}), already presented in Ref.~\cite{Nogga:2005hy},  
allows to take the limit $\lambda \rightarrow \infty$ analytically.  It remains to express
the states in Eq.~(\ref{eq:1.20}) in a more convenient fashion. Inserting the identity ${\bf
1}=G_0^{-1}(z) G_0(z)$ and using the representation $G(z)=G_0(z) +G_0(z) t(z)G_0(z)$ of the
full resolvent leads to 
\begin{eqnarray}
|\eta (z) \rangle &=& (1+t(z)G_0(z))|\phi\rangle \cr
 &=& (z-H_0) G(z) |\phi\rangle \cr
 &=& (z-H_0) \frac{1}{z-E_b} |\phi\rangle
\label{eq:1.21}
  \end{eqnarray}
Similarly, one obtains
\begin{eqnarray}
\langle \bar{\eta}(z) | &=& \langle \phi| (1+G_0(z)t(z)) \cr
  &=& \frac{1}{z-E_b}  \langle \phi| (z-H_0) \; .
 \label{eq:1.22}
\end{eqnarray}
Thus, the modified transition amplitude, in which the Pauli-forbidden state is projected to
infinity becomes with $z\equiv E$,
\begin{equation}
\tilde{t}(E) = t(E) - (E-H_0) \frac{|\phi \rangle \langle \phi |}{E-E_b} (E-H_0).
\label{eq:1.23}
\end{equation}
This modified two-body transition amplitude can easily be implemented in the Faddeev
equations as written in Eq.~(\ref{eq:1.9}). Since different channels may have 
Pauli-forbidden states at  different energies, in general, one has $E_b \equiv E_b(i)$. 
Although in this section we have dropped the explicit mention of the index $i$, Pauli-forbidden states in
different subsystems can 
be implemented  without any problem.

In case there are several Pauli-forbidden states in a specific channel of a subsystem, it is straightforward 
to generalize Eq.~(\ref{eq:1.23}) to give
\begin{equation}
\tilde{t}(E) = t(E) - (E-H_0)\sum_b \frac{|\phi_b \rangle \langle \phi_b |}{E-E_{b}} (E-H_0),
\label{eq:1.24}
\end{equation}
where $b$ runs over the number of Pauli-forbidden states. 

Although the expression of Eq.~(\ref{eq:1.23})  was presented in Ref.~\cite{Lehman:1982zz},
it was not used in this form. Rather the Faddeev equations were modified to explicitly
accommodate the two-body bound state being projected out. In fact, due to the difference
$(E-H_0)$, where $E$ is the energy of the subsystem and thus depending on the spectator
momentum $q$, the expression of Eq.~(\ref{eq:1.23}) is not a priori separable in the
coordinates needed in the Eqs.~(\ref{eq:1.13}). 
This makes the task of incorporating the Pauli
projection into the separable expansion somewhat challenging.

 To proceed, we first recall the basic properties of
 the generalized EST separable representation scheme~\cite{Ernst:1974zza}. The EST separable potential in
any given partial wave has the form
\begin{eqnarray}
v^{sep}(p',p)= \sum\limits_{lm}\;h_{l}(p')\;\lambda_{lm}\;h_{m}(p),
\label{eq:1.24a}
\end{eqnarray}
where the form factors are given as the off-shell $t$~matrices
\begin{eqnarray}
h_l(p')\equiv t(p',p_l;E_l),
\label{eq:1.24b}
\end{eqnarray}
corresponding to the original potential $V$. 
The strength of the potential is represented by matrix elements $\lambda_{lm}$ which depend entirely on the form factors.
This implies that the potential $v^{sep}(p',p)$ is completely determined
by the choice of form factors. According to Eq.~(\ref{eq:1.24b}), the latter are uniquely specified by the EST support points $\{E_l,p_l\}$, where $E_l$ is a fixed energy and $p_l$ a fixed momentum. We shall refer to $E_l$ as the support energy and $p_l$ the support momentum hereafter. The momentum $p_l$ can either be on-shell or off-shell for positive values of $E_l$. For negative support energies, $p_l$ is always off-shell. The number of EST support points give the rank of the separable expansion as well as the upper bound for
the indices $l$ and $m$. If the potential  used to
compute $t(p',p_l;E_l)$ supports a bound state,
the latter will be present in the separable expansion. If this bound state is a Pauli forbidden state, we
choose the potential $\tilde V$ defined in Eq.~(\ref{eq:1.14}) as the starting point in the EST
construction and  the Pauli forbidden state is projected out. 
Constructing a separable expansion of $\tilde V$ implies
that the form factors $h_l=t(p',p_l;E_l)$ in Eq.~(\ref{eq:1.24a})  are replaced by 
$\tilde h_l(p')= \tilde t(p',p_l;E_l)$. 
Additionally, the matrix elements $\lambda_{lm}$ must be replaced by $\tilde\lambda_{lm}$, where the latter are computed using the form factors $\tilde h_l(p')$.
Starting from Eq.~(\ref{eq:1.23})  the expression for the modified form factors is given by
\begin{equation}
\tilde{h}_{l}(p')  = h_l(p') - \frac{(E_l -E_{p'})}{(E_b-E_{p'})} h_b(p') \frac{1}{E_l
-E_b} h_b(p_l) 
    \frac{(E_l -E_{p_l})}{(E_b-E_{p_l})},
\label{eq:1.25}
\end{equation}
 where $h_b(p')\equiv \langle p'|V|\phi\rangle$. The momentum subscripts on the energy variables imply $E_{p_l}= 2\mu p_l^2$. 
 The explicit derivation of Eq.~(\ref{eq:1.25}) is given in Appendix~\ref{appendixB}. Using $\tilde h(p)$ in the separable expansion is straightforward and does not increase the rank. Multiple Pauli forbidden bound states simply produce additional modifications to the form factors in accordance with Eq.~(\ref{eq:1.24}).


\section{Results and Discussion}

\subsection{Interactions in the two-body subsystems}
\label{subsystems}

For computing the ground state of $^6$Li in a three-body model, we need the interactions in the 
different subsystems, $np$, $n\alpha$, and $p\alpha$.  For the $np$ subsystem, we employ the
CD-Bonn potential~\cite{Machleidt:2000ge} and include only the deuteron channel ($^3S_1-^3D_1$). This potential is one of the so-called
`high-precision' potentials that fit the two-nucleon observables up to $300$ MeV with  
$\chi^2\approx 1$. The proton and neutron masses given in Ref.~\cite{Machleidt:2000ge} are $m_p$~=~938.2723~MeV and 
$m_n$~=~939.5656~MeV.
For the  $n\alpha$~($p\alpha$) subsystem, we ignore the microscopic structure of the 
alpha-particle and employ a phenomenological interaction that is fitted to the
low-energy nucleon-alpha phase shifts. Here we include the $S_{1/2}$, $P_{1/2}$, and $P_{3/2}$ partial waves.
Our choice is the interaction given by
Bang~\cite{Bang:1979ihm}, which is of Woods-Saxon type, and supports a 
Pauli-forbidden  S-wave bound state.  
For this work, we slightly modify the Bang potential by  changing the central 
potential depth from $-43$~MeV to $-44$~MeV, to improve the description of the $n\alpha$ and $p\alpha$ 
phase shifts, particularly of the P-waves below $E_{lab}=10$~MeV.
As mass of the alpha particle we use $m_\alpha$~=~3727.379~MeV.
The S- and P-wave phase-shifts for $n+\alpha$ and $p+\alpha$ scattering
 calculated with this potential are shown in Fig.~\ref{fig1}.
They are compared to the phase shifts~\cite{Stammbach72} extracted from an R-matrix 
fit to data. 

To describe the $p\alpha$ interaction, we add to the $n\alpha$ potential a Coulomb force that
consists of a short-range part, corresponding to a charged sphere of radius $R_c = 1.25
\times 4^{1/3}$~fm  and the standard long-range point
Coulomb force ~\cite{Schellingerhout:1993ku}, 
\begin{equation}
V_c(r) =
  \begin{cases}
   \frac{Z_1Z_2e^2}{2R_c}\Big[ 3- \big(\frac{r}{R_c}\big)^2\Big]      & \quad r<R_c\\
    \frac{Z_1Z_2e^2}{r}  & \quad r>R_c ,\\
  \end{cases}
\label{eq:Coulomb}
\end{equation}
where $Z_1Z_2=2$, and $e^2$~=~1.43997~MeVfm.

\subsection{Binding energy of $^6$Li: separable vs non-separable}
\label{separable}

In this section, we consider two approaches to solve the momentum-space
Faddeev equations for the ground state of  $^6$Li using the two-body interactions described
in Section~\ref{subsystems} as input. 
 The first approach consists of solving
the bound state Faddeev equations directly as given by Eqs.~(\ref{eq:1.9}) leading to an `exact' solution
of the three-body bound state problem. The numerical results are obtained  using  Gauss-Legendre
quadratures. The momentum grids for converged results consist of $N_p$~=~200 points for the
pair momentum $p$ and $N_q$~=~200 for the spectator momentum $q$.
The maximum values for the above-mentioned momenta are set to  $p=60$~fm$^{-1}$ and $q=60$~fm$^{-1}$,
respectively.
This calculation
yields  $E_{3}=-3.787\text{~MeV}$ for the three-body binding energy of $^6$Li  when no Coulomb
interaction is included and
$E_{3}=-2.777\text{~MeV}$ with the Coulomb interaction of Eq.~(\ref{eq:Coulomb}).
The Coulomb potential is treated by introducing a cutoff radius $R_{cut}$ beyond which $V_c(r)$ is set to zero. The momentum space representation is evaluated using either an analytic or numerical Fourier transform. Both methods are numerically stable. To further test the numerical
stability of the calculation, the binding energy was computed using different values
of the cutoff radius. We found that the result for $E_3$ is independent of the cutoff radius for $R_{cut}>$~15~fm.
 The experimental value is  $E^{exp}_{3}=-3.699\text{~MeV}$ from Ref.~\cite{AjzenbergSelove:1988ec}.
Our three-body calculation slightly underbinds $^6$Li, a standard feature of these three-body models. The
difference is typically accounted through a three-body interaction~\cite{Zhukov:1993aw}.

The second approach for solving the Faddeev equations consists of two steps. First, the EST~\cite{Ernst:1974zza}
scheme is employed to construct separable representations of
the two-body potentials given in Section~\ref{subsystems}.
Then, the separable interactions are used to
solve Eqs.~(\ref{eq:1.13}) in order to obtain the three-body binding energy as well as the Faddeev amplitude according to Eq.~(\ref{eq:1.12}).
In the current example, the separable expansion is used to make a
prediction for the $^{6}$Li three-body binding energy with
a precision of four significant figures. To check the accuracy of this
prediction, the results are compared to the ones obtained directly
without the separable expansion.

According to Eq.~(\ref{eq:1.24b}), the  EST separable expansion employs solutions of the LSE as basis states.
These states depend on two parameters, the two-body
energy $E_l$ as well as the asymptotic momentum $p_l$. We refer to each combination of $E_l$ and $p_l$ as an EST support point.
It should be pointed out that if one employs the constraint $E_l=2\mu{p}_l^2$, with $\mu$ being the
reduced mass of the two-body system, the basis states depend only on one parameter.
While the EST scheme~\cite{Haidenbauer:1986zza,Cornelius:1990zz,Nemoto:1998wt}
has been applied in solving Faddeev equations in separable form, those works did not take
advantage of the full parameter space by imposing  $p_l=\sqrt{2\mu |E_l|}$.
We make use of the full parameter space for the basis states and choose
$p_l$ and $E_l$ independently as suggested in Ref.~\cite{Ernst:1974zza}.
The bound state Faddeev equations require off-shell two-body $t$-matrices as
input in the energy range $-\infty \le E_{2b}\le E_{3}$.
Therefore, a good separable representation of the off-shell properties of the $t$-matrices is
required to reproduce the direct calculation accurately.

A successful application of the EST scheme hinges on
an effective selection of the support energies and momenta.
Since we are interested in a separable expansion for two-body energies $E_{2b}$ between $-\infty$ and $E_{3}$, we
restrict ourselves to negative energy support points.
The off-shell $t$~matrix has a smooth energy-dependence and
is dominated by the energy-independent Born term at large values of $|E_{2b}|$.
It is thus not necessary to incorporate support points at large negative energies. In practice, it is sufficient to consider support energies in
the range $-100~\text{MeV}\le E_l\le 0$.

The use of separable expansions for the two-body potentials introduces
uncertainty in three-body observables. This uncertainty
must be quantified in order to make meaningful predictions.
The  dependence of $E_3$ on the choice of support points reflects the uncertainty in our procedure.
 By varying the latter while keeping the rank fixed
 can lead to a quantitative estimate of this uncertainty.
Carrying out this procedure for successively increasing ranks provides means
for making precise predictions of three-body
observables using this approach.

Contrary to the smooth energy dependence of the off-shell $t$~matrix, its
dependence on the off-shell momenta is much more intricate and is determined
by the shape of the underlying potential. As a consequence, the
predicted three-body observables show more sensitivity to the choice of the
support momenta. To make an economic choice for the latter, we first identify
the relevant range of the $t$~matrix in momentum space and define the support momenta within it.

To illustrate how the support momenta are chosen, the off-shell
$t$~matrix corresponding to the CD-Bonn potential is computed for
the $^3S_1-^3D_1$ partial wave. Figure~\ref{fig:3.2.1}
shows the off-shell $t$~matrix elements $t_{{l_{np}}'l_{np}}~(p',p;E_{2b})$ for the $np$ system as a function of the off-shell momentum $p$. The matrix elements $t_{00}~(p',p;E_{2b})$ are depicted in panel (a) while $t_{22}~(p',p;E)$ and $t_{20}~(p',p;E)$ are shown in
panels (b) and (c). The center-of-mass energy (c.m.) is fixed at $E_{2b}=-50$~MeV. Results
obtained using the CD-Bonn potential are indicated by solid lines
for $p'=0.3$~fm$^{-1}$ and dashed lines for   $p'=0.8$~fm$^{-1}$. Corresponding
$t$~matrix elements calculated using a rank-6 separable representation of the CD-Bonn potential are illustrated by  triangles for $p'=0.3$~fm$^{-1}$ and diamonds for   $p'=0.8$~fm$^{-1}$. The
energies are in units of MeV while the momenta are given in fm$^{-1}$. The
support points are $\{E_l,p_l\}~=\{-60,0.4\},\{-60,1.1\},\{-60,2.5\},\{-5,0.4\} ,\{-5,1.1\},\{-5,2.5\}$.
As mentioned above, the support energies are selected within the range
$-100~\text{MeV} \le E_l\le 0$. Their specific values  can be altered
without compromising the accuracy of the separable representation. However,
the support momenta are  chosen to reproduce the structure of the
$t$~matrix below 5~fm$^{-1}$. As a first guess, the momenta  are chosen such that there is one in the
vicinity of each minimum or maximum. Improvement of the separable expansion is attained by further adjustment of the initial values.

The choice of momenta is not unique since
a slight change in the given values can still capture the structure of the
off-shell $t$~matrix. However, changing the value of each support momentum
by, e.g. 0.5~fm$^{-1}$, can
already lead to a poor representation of the $t$~matrix,  as well as the three-body
observables. It is thus imperative to check
that each chosen set of momenta captures
the shape of the off-shell $t$~matrix in order to ensure that the separable expansion converges rapidly. It should be noted that, although the structure of the $t$~matrix differs for each $E$ and $p'$, the regions of
intricate momentum dependence remain mostly unaltered. For example, this can
be seen by comparing the $t$~matrix at $p'=0.3$~fm$^{-1}$ and $p'=0.8$~fm$^{-1}$. Although the shape is quite different in each case, the features that determine the location of the support momenta are situated at similar positions. Consequently, the support points adjusted to reproduce the off-shell $t$~matrix at $p'=0.3$~fm$^{-1}$
are  equally well suited for $p'=0.8$~fm$^{-1}$. Thus, by accurately representing the off-shell $t$~matrix at a single energy by including several support momenta, one can obtain an accurate representation of the off-shell $t$~matrix at other energies. Although such a choice is specific to the CD-Bonn potential,
these support points would be applicable to any $NN$ $t$~matrix that exhibits either (1) a similar off-shell structure or (2) a considerably less complicated dependence on the off-shell momenta. The structure of the
$NN$ $t$~matrix corresponding to most high precision and chiral potentials
is similar in the low momentum region that determines the support momenta.
We thus expect that the support points determined for the CD-Bonn potential
will provide an equally good representation for all such $NN$ potentials.
For example, we verified that those same support points yield excellent results for the high precision Nijmegen~I~\cite{Stoks:1994wp} and AV18~\cite{Wiringa:1994wb} as well as the chiral potential
of Ref.~\cite{Entem:2003ft}. Contrarily, the structure of the off-shell    
$t$~matrix corresponding to the Woods-Saxon Bang potential is very different from that of the $NN$ $t$~matrices,
and thus an independent determination of the support momenta must be carried out.

To quantify the uncertainty on the three-body binding energy, separable representations of successively increasing
rank are constructed for both,  the CD-Bonn and the Bang potential.
Table~\ref{table:est1} shows several
separable representations of the CD-Bonn potential. The first and
second columns give the label and rank of the separable potential.
The EST support energies and momenta are listed in the third
and forth columns, respectively. The same information is given in
Table~\ref{table:est2} for the Bang potential. To proceed, we first fix the EST support points
for the Bang interaction while varying those of the CD-Bonn potential.
Table~\ref{table:eb1} shows the three-body binding energies for the ground state of $^6$Li
calculated using a variety of $np$ separable representations taken from
Table~\ref{table:est1}. For this study we do not include the Coulomb interaction.
 The EST8-4 separable representation of the Bang
interaction defined in Table~\ref{table:est2} is adopted in the $n\alpha$ and $p\alpha$ subsystems.
To ease comparison, we include in the last rows of Tables~\ref{table:eb1}
and~\ref{table:eb2}  the exact results obtained when solving  Eq.~(\ref{eq:1.9}) directly.

We observe that the numerical value for the binding energy fluctuates
as the support points are varied. However, the fluctuations decrease
as the rank of the separable potentials is increased. From
Table~\ref{table:eb1} we see that the uncertainty in the binding
energy is $\delta E_{3}\approx 50$~keV for the rank~3 representation. Increasing
the rank to five reduces the uncertainty down to $\delta E_{3}\approx 5$~keV.
A further increase of the rank to six reduces the uncertainty to $ \delta E_{3}\approx 0.5$~keV, which corresponds to a precision of four significant figures. In addition to the uncertainty associated
with the selection of the support energies, one must take into account
the convergence of the binding energy with respect to the rank
of the separable potential. From Table~\ref{table:eb1}
we see that increasing the rank from six to seven leaves the fourth
digit of the binding energy  unaltered. This observation, together
with the fact that $ \delta E_{3}< 0.5$~keV,  guarantees that the
numerical result for $E_3$ is precise to four significant figures.

Next, the support points for the CD-Bonn potential are fixed and those
corresponding to the Bang potential are varied.
Table~\ref{table:eb2} is the same as Table~\ref{table:eb1} but shows
results for different separable expansions of the Bang interaction. The EST8-1 separable representation of the CD-Bonn
potential taken from Table~\ref{table:est1} is adopted for
the $np$ subsystem.
Here we observe that rank-3 and rank-4 potentials lead to the
uncertainties $\delta E_{3b}\approx 40$~keV  and $\delta E_{3b}\approx 13$~keV.
Moreover, a rank-7 representation is needed in order to obtain
an uncertainty of approximately 0.5~keV. To ensure that the binding energy is converged
to at least four significant figures, it is necessary to increase the
rank to 8. The predicted value for the three-body binding energy can thus be read off from Tables~\ref{table:eb1} and~\ref{table:eb2} as $E_{3b}=-3.787$~MeV, in perfect agreement with the exact result.

The rapid reduction of the uncertainty observed in Tables~\ref{table:eb1} and~\ref{table:eb2}  is primarily due to the efficient choice of the support momenta. To illustrate this point,
we consider calculations in which the
constraint $p_l=\sqrt{2\mu |E_l|}$ is imposed. We choose three sets of support
support energies for the CD-Bonn potential, namely, $E_l=[-150,-120,-80,-60 ,-45,$
$-35,-15,-5\}$~MeV, $\{-180,-140,-100,-70,-55,-35,-10,-3\}$~MeV, and
$\{-200,-160,-120,-80,-40,-25,-10,$
$-4\}$~MeV. These sets yield $E_{3b}=-3.803$~MeV, $E_{3b}=-3.788$~MeV, and $E_{3b}=-3.795$~MeV, respectively. Here we see that despite being rank-8, these representations lead to fluctuations in the third digit.
This demonstrates that, in order to obtain a result that is precise to four significant figures, it is
essential that the full parameter space for choosing a basis is considered and
 the support momenta are chosen independently from the support energies.

Finally we calculate the three-body binding energy of $^6$Li  when the Coulomb interaction of
Eq.~(\ref{eq:Coulomb})
is included in the description of the $p\alpha$ subsystem. The Coulomb interaction leads to a different
structure of the
 the $p\alpha$ potential, and the above analysis has to be repeated,
leading to a different
set of support points. The rank required to obtain a precision
of at least four significant figures remains unchanged at eight.
Using a rank-8 separable representations for the Coulomb and
Bang potentials yields a three-body binding energy of -2.777~MeV
which agrees completely with the exact calculation.
The support points were chosen to be
$\{E_l,\;p_l\}$~=~\{-55,0.2\}, \{-55,1.0\}, \{-55,1.2\}, \{-55,3.0\}, \{-3,0.2\}, \{-3,1.0\}, \{-3,1.2\}, \{-3,3.0\}.

Lastly, we want to comment that it is mandatory to perform the projection procedure for the Pauli-forbidden state in the separable representation
as in the exact calculation. Creating a separable representation that attempts to exclude the Pauli-forbidden state a priori leads to erroneous results.

\subsection{Properties of $^6$Li}
\label{properties}

After discussing the convergence and accuracy of the three body binding energy of $^6$Li,
we need to consider
properties of the wave function obtained in both schemes, since we do not only want to have
excellent agreement in the three-body binding energy  but also in observables derived from the wave
function.  To this aim, we consider the momentum distributions with respect to the
Jacobi coordinates of the wave function $\Psi(\vec{p},\vec{q})$. Choosing a specific set of
Jacobi variables, e.g. the set $(ij)k$, in which $k$ is the spectator with respect to the
pair $(ij)$, the momentum distribution of the spectator is given as
\begin{equation}
n(q_k)=\sum_{\alpha_k} \int dp_k p_k^2 | \Psi_{\alpha_k}(p_k,q_k)|^2,
\label{eq:nq}
\end{equation}
and the momentum distribution of the pair is given as
\begin{equation}
n(p_k)=\sum_{\alpha_k} \int dq_k q_k^2 | \Psi_{\alpha_k}(p_k,q_k)|^2.
\label{eq:np}
\end{equation}

The momentum distribution of the different pairs in the ground state of $^6$Li are
shown in panels (a) and (b) of  Fig.~\ref{fig2} on a linear as well as a logarithmic scale, where
the corresponding pair is indicated in the round brackets of the legend. The solid, dashed as
well as dotted lines are calculated using the non-separable forces, whereas the crosses,
downward and upward triangles correspond to the same calculation using separable forces. The calculations are in excellent agreement. For small
momenta, the distribution in the $(np)$ pair is about twice as large as the ones in the
$(n\alpha)$ and $(p\alpha)$ pairs, whereas for momenta larger than 2~fm$^{-1}$ there is an
order of magnitude (or more) difference between the momentum distribution in the $(np)$ pair
and the $(n\alpha)$ and $(p\alpha)$ pairs, an indication of the high momentum components of the CD-Bonn potential.

Panels (c) and (d) of Fig.~\ref{fig2} depict the momentum distributions of the spectator
particle with respect the pair given in brackets in the legend. For very small momenta $q$, the
distribution of the alpha momentum with respect to the $(np)$ pair dominates by an order of
magnitude over the ones of the two other spectator momenta. However, the logarithmic scale in
panel (d) shows that for different values of $q$, these roles interchange twice. Finally, for
$q \ge 3.5$~fm$^{-1}$ the distributions in which either the proton or the neutron are the
spectators dominate, which, again is a reflection of the high momentum components of the CD-Bonn  potential.

As discussed in Sec.~\ref{subsystems}, the effective interaction between the neutron (proton)
and the alpha particle is represented by Woods-Saxon type potentials. Thus, in both
subsystems there is a bound state in the S$_{1/2}$-state, for the $n\alpha$ subsystem this
bound state is at -10.326~MeV and for the $p\alpha$ subsystem at -8.879~MeV. Those bound
states are forbidden by the Pauli principle, and need to be projected out using the
formulation outlined in Sec.~\ref{pauliblocking}. In both subsystems we introduce
an additional term to the potential according to Eq.~(\ref{eq:1.14}),
 $\hat{V}^i = \lambda_i |\phi^i\rangle \langle \phi^i|$, and let the parameters $\lambda_i$ go to infinity.

In order to better understand the action of the parameters
$\lambda_{n\alpha}=\lambda_{p\alpha} \equiv \lambda$,
we choose a set of finite values for $\lambda$ and calculate the ground state
three-body binding energy
and expectation value as function of $\lambda$. The results of these calculations
are listed in Table~\ref{table-lam}. To simplify this study,  the Coulomb potential is here
omitted, leading to $V_{n\alpha}=V_{p\alpha}$.
The expectation value of the total Hamiltonian is in this case given by
\begin{eqnarray}
\langle E_3(\lambda) \rangle &\equiv &\langle \Psi(\lambda) | H_{3b} | \Psi(\lambda) \rangle \cr
&= & \langle \Psi(\lambda) | H_0 + V_{np} +  2 {V}_{n\alpha} | \Psi(\lambda) \rangle,
\label{eq:expect}
\end{eqnarray}
where $H_{3b}$ is the three-body Hamiltonian.
The values of $E_3 (\lambda)$ obtained from the solution of the Faddeev equation, Eq.~(9), start to
agree with the expectation value calculated using Eq.~(\ref{eq:expect}) within 4 significant
figures once $\lambda$ exceeds 1000~fm$^{-1}$. Letting $\lambda \rightarrow \infty$ gives
perfect agreement. In order to illustrate that the Pauli forbidden  S$_{1/2}$ state
$|\phi_{n\alpha}\rangle$  is completely projected
out for $\lambda \rightarrow \infty$, we define a probability
\begin{equation}
\mathcal{P}_{TDB}(\lambda)=\sum_{\alpha_k'}\int_0^\infty q_k^2 \Bigg|\int_0^\infty d{p_k'}\;{p'_{k}}^2\;
\phi_{\alpha_k'}(p_{k}')\Psi_{\alpha_k'}(p'_k,q_k;\lambda)\Bigg|^2 dq_k ,
\label{eq:PFstate}
\end{equation}
which gives the overlap between the Pauli forbidden  S$_{1/2}$ state and the $^6$Li
ground state wave function calculated for a specific $\lambda$.
A detailed discussion of this probability is provided in Appendix~\ref{appendixc}.
Obviously, this quantity is calculated in the Jacobi coordinates where  $n\alpha$
constitutes the subsystem.  The calculated values of $\mathcal{P}_{TDB}(\lambda)$
are listed in the last column of Table~\ref{table-lam} and clearly indicate that
for $\lambda \ge 10^4$~fm$^{-1}$ the overlap is numerically zero.

Studying the evolution of the three-body binding energy as function of the parameter $\lambda$ shows how
the deep three-body bound state including the Pauli forbidden states in the $n\alpha$ and
$p\alpha$ subsystems moves to the physical three-body bound state. However, the binding
energy does not give further information about the characteristics of the bound state.
Since the Pauli forbidden states in the $n\alpha$ and
$p\alpha$ subsystems occur in the S$_{1/2}$ partial wave, the unphysical deep bound
state should be dominated by this partial wave. However, we know that
the physical ground state is dominated by P$_{3/2}$ components.
It is thus illustrative to investigate how the components of the ground state
wave function change as a function of the parameter $\lambda$.
To proceed, we note that the probability for each partial wave state $|\alpha_k\rangle$  in the three body
wavefunction is given by
\begin{equation}
\langle \Psi_{\alpha_k}(\lambda)  | \Psi_{\alpha_k}(\lambda)  \rangle = \int\limits_{0}^{\infty} dp_k\; dq_k\; p_k^2 q_k^2\;
    \big|\Psi_{\alpha_k}(p_k,q_k;\lambda)\big|^2.
\label{eq:probability}
\end{equation}
Here the index $k$ represents the Jacobi coordinate $(n\alpha)p$ in which the proton is the spectator with momentum $q_k$. 
We recall that the three-body angular momentum states $|\alpha_k\rangle$ are constructed by coupling angular momentum states of the pair $|\beta_k\rangle$ to those of the
spectator $|\gamma_k\rangle$, so that $|\alpha_k\rangle=|\beta_k\rangle\otimes|\gamma_k\rangle$. In the present case $|\beta_k\rangle$ corresponds to the $S_{1/2}$, $P_{1/2}$, and $P_{3/2}$ partial wave states of the $n\alpha$ subsystem. 
To determine the probability  for each of those two-body states, one must sum over the angular momenta of the spectator $\gamma_k$. The probability for a state  $|\beta_k\rangle$ is thus given by 
\begin{equation}
N_{\beta}(\lambda)=\sum\limits_{\gamma}\int\limits_{0}^{\infty} dp\; dq\; p^2 q^2\;
    \big|\Psi_{\alpha}(p,q;\lambda)\big|^2,
\end{equation}
where $\alpha=\{\beta,\gamma\}$ and the subscript $k$ is omitted for concision. Figure~\ref{fig3} shows the values of $N_{\beta}(\lambda)$ as a function of $\lambda$ for the $^6$Li three-body ground state.  The solid, dashed, and dot-dashed lines represent the $S_{1/2}$, $P_{3/2}$ and $P_{1/2}$ partial wave states of the $n\alpha$ subsystem.
The vertical line indicates the value of $\lambda$ for which the $n\alpha$ system
becomes unbound.
As expected, for $\lambda=0$ the ground state is
completely dominated by the $S_{1/2}$ state. This remains true for values
of $\lambda$ smaller than $0.05$ fm$^{-1}$.  It is worthwhile to note that, even when the
$n\alpha$ subsystem becomes unbound, the three-body ground state
of $^6$Li is still dominated by the $S_{1/2}$ component. Only when $\lambda$ approaches
0.1~fm$^{-1}$, the probability of the $S_{1/2}$ component rapidly decreases. The 
corresponding probability of the $P_{3/2}$ rapidly increases to its final value of about 70\%.
Moreover, the ground state acquires a  $P_{1/2}$ probability of about 20\% and maintains an $S_{1/2}$
probability of about 10\% which is due to the continuum states of $n\alpha$, $p\alpha$, and $np$ subsystems.

\section{Summary and Outlook}
\label{summary}

In this work, we explore solving momentum-space Faddeev equations using separable interactions based on the EST scheme \cite{Hlophe:2013xca,Hlophe:2014xca,Hlophe:2015rqn}, for bound three-body systems of the type $n+p+A$. Our goal is to benchmark  this separable method against the standard approach of directly solving momentum-space Faddeev equations. 
We apply both approaches to $^6$Li, taking the CD-Bonn~\cite{Machleidt:2000ge}  interaction for the $np$ pair and the Bang~\cite{Bang:1979ihm}  potential for the $n$($p$)-$\alpha$ subsystems. Our results for the $^6$Li bound state demonstrate that using a
separable implementation of the Faddeev equations is equivalent to solving them directly: 
the binding energies obtained with the separable interactions agree within 4-digits with the exact calculation,
 and the  momentum distributions are also in perfect agreement. Our values of the binding energy obtained for $^6$Li with CD-Bonn and Bang are consistent with previous three-body calculations.  Since we are dealing with a bound state problem, including the Coulomb interaction in the momentum-space Faddeev equations does
not present a problem in both approaches.

As a consequence of our study, there are a few important developments worth highlighting. 
First and foremost, we extended the EST construction of the separable interaction to include off-shell 
properties of the t-matrix by allowing energy and momentum support points to be chosen independently. 
This proved to be critical for the high quality description of the properties
of the three-body system and to achieve the desired 4-digit precision. Second, the energy and momentum
support points developed for the $np$ subsystem are independent of the choice of the NN interaction as long 
as it describes the low energy behavior of the deuteron channel with high precision. 
The numerical implementation valid for the CD-Bonn interaction will transport immediately to other high
precision NN potentials as well as chiral NN potentials. However, when solving the $n+p+A$ problem for
bound systems where $A>4$, and given the wide range of nucleon-nucleus effective interactions available,
we expect one will need to inspect the properties of the two-body nucleon-nucleus t-matrices carefully
 and revisit the issue of optimum energy and momentum support points in those cases again. 
Similar to \cite{Hlophe:2013xca,Hlophe:2014xca,Hlophe:2015rqn}, we find here that the structure of 
the two-body t-matrices as a function of energy and momentum determines the minimal rank needed for 
an accurate description of both the two-body and three-body observables.
 
Another important development resulting from this study concerns the method used to project two-body 
Pauli-forbidden states out of the model space. 
We have developed an approach that does not modify the Faddeev equations and thus can be
implemented straightforwardly in momentum-space Faddeev equations 
either in their non-separable or  EST-type separable representation. 
This approach is effective in projecting out the forbidden state at a minimal computational cost. 
We also provide a generalization for dealing with an arbitrary number of Pauli-forbidden states 
in a computationally efficient manner. This will be essential when moving to heavy systems.

This work lays the ground to now proceed to three-body scattering with EST-separable interactions. In the separable 
formulation, the Coulomb interaction can be accurately taken into account even for complex nuclei  with large $Z$ 
as outline in Ref.~\cite{Mukhamedzhanov:2012qv}. 
The next step is to tackle $d+A$ elastic scattering below  and above the three-body breakup threshold, 
followed by the ultimate goal of applying the method to deuteron induced nuclear reactions on heavy 
ions, at energies well above three-body breakup threshold.
â

\newpage

\appendix

\section{Explicit representation of non-separable and separable Faddeev equations}
\label{appendixA}

Here we summarize the explicit expressions entering our formulation of the Faddeev Eqs.~(\ref{eq:1.9}) and  (\ref{eq:1.13}). 
Besides the $t$-matrix, the Faddeev equations in non-separable form require coordinate transformations from Jacobi momenta 
that single out particle $i$ to Jacobi coordinates that single out particle $k$. These are most conveniently  performed 
separately for orbital and spin space. Therefore, the basis states of Eq.~(\ref{eq:1.7}) are first recoupled into an LS basis and 
then the transformation is applied. The resulting geometrical function is then
\begin{eqnarray}
{\cal G}_{\alpha_{k} \alpha_{i}} (p_{k}q_{k}x) & =& \sum_{LS} (2S+1) \sqrt{(2J_{ij}+1)(2J_{k}+1)(2J_{jk}+1)(2J_{i}+1)}
\left\{  \begin{array}{ccc}   l_{k}  & s_{k} & J_{ij}  \cr 
\lambda_{k} & j_{k}  & J_{k} \cr 
L & S & J 
\end{array}\right\} \, 
\left\{  \begin{array}{ccc}   l_{i}  & s_{i} & J_{jk}  \cr 
\lambda_{i} & j_{i}  & J_{i} \cr 
L & S & J 
\end{array}\right\}
 \nonumber \\
&& \times {8\pi^{2}} \, \sum_{M=-L}^{L} \left\{ Y_{l_{k}}^{*}(\hat p_{k}) \, Y^{*}_{\lambda_{k}}(\hat q_{k})   \right\}^{LM}  
\left\{ Y_{l_{i}}(\reallywidehat {-\alpha \vec p_{k} - \beta \vec q_{k} }) \, Y_{\lambda_{i}}( \reallywidehat {\vec p_{k} - \gamma \vec q_{k} } )   \right\}^{LM} 
\nonumber \\
& & \times  (-)^{s_{i}+2j_{i}+j_{j}+j_{k}} \sqrt{(2s_{k}+1)(2s_{i}+1)}  \left\{  \begin{array}{ccc}   j_{i}  & j_{j} & s_{k}  \cr 
j_{k} & S  & s_{i} \cr 
\end{array}\right\} 
 \ \ . 
\end{eqnarray}

The spherical harmonics $Y_{lm}(\hat p)$ dependent on the angles $\hat p$ of the  vector $\vec p$. For the evaluation, we choose a coordinate 
system where the pair momentum is angular independent and the spectator momentum  is in the $x$-$y$ plane: 
\begin{equation}
\vec p_{k} = \left( \begin{array}{c}
0 \cr 0 \cr p_{k}
\end{array}\right) 
\hspace{1cm} \vec q_{k} = \left( \begin{array}{c}
q_{k} \sqrt{1-x^{2}} \cr 0 \cr q_{k} x 
\end{array}\right)  \ \ .
\end{equation}
The curly brackets grouping the spherical harmonics indicate that they are coupled to a state of total orbital 
angular momentum $L$ and third component $M$.  
The mass ratios are given by  
\begin{eqnarray}
\alpha & =& \frac{m_{k}}{m_{j}+m_{k}} \\ \nonumber
\beta  & = & \frac{(m_{i}+m_{j}+m_{k})\, m_{j}}{(m_{i}+m_{j})(m_{j}+m_{k})} \\ \nonumber
\gamma & =& \frac{m_{i}}{m_{i}+m_{j}} \ \ .
\end{eqnarray}
For this case, the shifted momenta are given by 
\begin{eqnarray}
\pi'_{i}(p_{k}q_{k}x) & =& \sqrt{\alpha^{2} p_{k}^{2} + \beta^{2} q_{k}^{2} + 2 \alpha \beta p_{k}q_{k}x}  \\ \nonumber
\chi'_{i}(p_{k}q_{k}x) & =& \sqrt{p_{k}^{2} + \gamma^{2} q_{k}^{2} - 2 \gamma p_{k}q_{k}x}  \ \ .
\end{eqnarray}
For the derivation, we followed similar steps as in Ref.~\cite{Balian:1969sd}. Different but equivalent 
expressions that involve Legendre polynomials can be derived \cite{Glockle:1983un}. We used the ones given above 
since the numerical implementation is stable even for large orbital angular momenta. 

For the derivation of Eq.~(\ref{eq:1.13}), we insert the separable expansion Eq.~(\ref{eq:1.12}) into the Faddeev equations.
It is then advantageous to substitute the $p_{k}'$ integral by an integral over $q_{i}$ or $q_{j}$, respectively. 
The Faddeev equations then read 
\begin{eqnarray}
\psi_{k}^{\alpha_k}(p_k, q_k) &=&  G_0(E_{q_k},p_k) 
   \sum_{\alpha_k'} \int_{-1}^{1} dx \,  \Bigg[ \int dq_i q^2_i t^{\alpha_k \alpha'_k}_{k} (p_k, \pi_{k};E_{q_k})
  \sum_{\alpha_i'} \tilde {\cal G}_{\alpha'_k \alpha'_i}(q_k q_{i} x) \psi_{i}^{\alpha'_i} (\pi'_i,q_i) \\ \nonumber 
   & & \hspace{3.2cm} + \int dq_j q^2_j t^{\alpha_k \alpha'_k}_{k} (p_k, \pi_k ;E_{q_k}) \sum_{\alpha_j'} \tilde {\cal G}_{\alpha'_k \alpha'_j}(q_k q_{j} x) \psi_{j}^{\alpha'_j} (\pi'_j, q_j)\Bigg] \ \ .
\label{eq:1.9a}
\end{eqnarray}
In this case, the geometrical function is  defined by 
\begin{eqnarray}
\tilde {\cal G}_{\alpha_{k} \alpha_{i}} ( q_{k}q_{i}x) & =& \sum_{LS} (2S+1) \sqrt{(2J_{ij}+1)(2J_{k}+1)(2J_{jk}+1)(2J_{i}+1)}
\left\{  \begin{array}{ccc}   l_{k}  & s_{k} & J_{ij}  \cr 
\lambda_{k} & j_{k}  & J_{k} \cr 
L & S & J 
\end{array}\right\} \, 
\left\{  \begin{array}{ccc}   l_{i}  & s_{i} & J_{jk}  \cr 
\lambda_{i} & j_{i}  & J_{i} \cr 
L & S & J 
\end{array}\right\}
 \nonumber \\
&& \times 8\pi^{2} \, \sum_{M=-L}^{L} \left\{ Y_{l_{k}}^{*}(\reallywidehat {\gamma \vec q_{k} +  \vec q_{i} }) \, Y^{*}_{\lambda_{k}}(\hat q_{k})   \right\}^{LM}  
\left\{ Y_{l_{i}}(\reallywidehat {-\vec q_{k} -  \alpha \vec q_{i} }) \, Y_{\lambda_{i}}( \hat q_{i} )   \right\}^{LM} 
\nonumber \\
& & \times  (-)^{s_{i}+2j_{i}+j_{j}+j_{k}} \sqrt{(2s_{k}+1)(2s_{i}+1)}  \left\{  \begin{array}{ccc}   j_{i}  & j_{j} & s_{k}  \cr 
j_{k} & S  & s_{i} \cr 
\end{array}\right\} 
 \ \ . 
\end{eqnarray}
The momentum vectors are chosen as 
\begin{equation}
\vec q_{k} = \left( \begin{array}{c}
0 \cr 0 \cr q_{k}
\end{array}\right) 
\hspace{1cm} \vec q_{i} = \left( \begin{array}{c}
q_{i} \sqrt{1-x^{2}} \cr 0 \cr q_{i} x 
\end{array}\right),  
\end{equation}
and the shifted momenta change to  
\begin{eqnarray}
\pi_{k}(q_{k}q_{i}x) & =& \sqrt{\gamma^{2} q_{k}^{2} + q_{i}^{2} + 2 \gamma q_{k}q_{i}x},  \\ \nonumber
\pi'_{i}(q_{k}q_{i}x) & =& \sqrt{q_{k}^{2} + \alpha^{2} q_{i}^{2} + 2 \alpha q_{k}q_{i}x}.
\end{eqnarray}
Using this form of the Faddeev equations, it is easy to read off Eq.~(\ref{eq:1.13}). Since the form factors 
of the separable interaction are given a priori, it is possible to precalculate the angular integral 
leading to the definition
\begin{equation}
Z^{(ki)}_{n\alpha_{k},\nu\alpha_{i}}(q_k, q_i) = \int_{-1}^{1} dx \ h_n^{\alpha_k}(\pi_k) \tilde {\cal G}_{\alpha_k \alpha_i}(q_k q_{i} x) G_0(E_{q_i},\pi'_{i}) h_\nu^{\alpha_i}(\pi'_i) \ \ .
\end{equation}

The wave functions cannot be represented in a separable form. They are obtained from  $\psi_{k}^{\alpha_k}(p_k, q_k)$ using 
Eq.(\ref{eq:1.2}). Thereby further coordinate transformations using
either  ${\cal G}_{\alpha_{k} \alpha_{i}} (p_{k}q_{k}x)$ or 
$\tilde {\cal G}_{\alpha_k \alpha_i}(q_k q_{i} x)$  are required to represent all three Faddeev components in the same set of coordinates.

\section{Projecting Pauli-forbidden states in case of separable potentials}
\label{appendixB}
In order to set up the formulation for projecting a Pauli-forbidden state to infinity when
using separable potentials based in the EST formulation, let us have a closer look at the
functions $|  \eta(z)\rangle$ and $\langle \bar{\eta}(z)|$ of Eqs.~({\ref{eq:1.20}).
The explicit momentum space representation reads
\begin{eqnarray}
\langle \bar{\eta}(z)|p\rangle &=& \langle \phi |\big[ 1+G_0(z)t(z)\big] |p\rangle \cr
 &=& \langle \phi | G(z) G_0(z)^{-1}|p\rangle \cr
 &=& \frac{z-E_p}{z-E_b} \langle \phi| p\rangle \cr
&=& \frac{z-E_p}{z-E_b} \langle \phi| V G_0(E_b)| p\rangle\cr
&=& \frac{z-E_p}{z-E_b} \langle \phi| V|p\rangle \frac{1}{E_b - E_p},
\label{eq:b.1}
\end{eqnarray}
where we used the Schr\"odinger equation for $\langle \phi|p\rangle$.
The momentum subscript, $p$, on the energy variable, $E$, implies $E_{p}=p^2/2\mu$,
while the bound state energy is represented by $E_{b}$. 
Similarly, one obtains
\begin{eqnarray}
\langle p'|\eta(z)\rangle &=& \langle p'| \big[1+t(z)G_0(z)\big]|\phi\rangle \cr
&=& \frac{1}{E_b-E_{p'}} \langle p'|V|\phi\rangle \frac{z-E_{p'}}{z-E_b} \; .
\label{eq:b.2}
\end{eqnarray}
Putting everything together and setting $z \equiv E$, Eq.~(\ref{eq:1.23}) takes the explicit form
\begin{equation}
\tilde{t}(p',p;E)= t(p',p;E) - \frac{(E-E_{p'})}{(E_b-E_{p'})} \; \frac{(E-E_p)}{(E_b-E_p)} \; 
    \frac{\langle p'|V|\phi\rangle \langle \phi|V|p\rangle}{E-E_b}.
\label{eq:b.3}
\end{equation}
The above expression clearly shows that the half-shell elements of $\tilde{t}(p',p;E)$ and
$t(p',p;E)$ are identical, and for $E \rightarrow E_b$ the pole of $t$ is removed for $\tilde t$.
However, due to the differences $(E-E_{p'})$ and $(E-E_{p})$, and having
in mind that energy $E$ of the subsystem depends on the spectator momentum $q$ in the Faddeev
equations, $\tilde{t}(p',p;E)$ is not yet in the separable form needed in
Eqs.~(\ref{eq:1.13}). 
At a specific off-shell support energy $E_l$ and support momentum $p_l$ we have 
\begin{equation}
\tilde{t}(p',p_l;E_l)= t(p',p_l;E_l) - \frac{(E_l-E_{p'})}{(E_b-E_{p'})} \;
\frac{\langle p'|V|\phi\rangle \langle \phi|V|p_l\rangle}{E_l-E_b} \;
\frac{(E_l-E_{p_l)}}{(E_b-E_{p_l)}},
\label{eq:b.4}
\end{equation}
having in mind that these off-shell EST functions are characterized by $\{E_l,p_l\}$, and only
need to be linearly independent and solutions of a Lippmann-Schwinger type integral
equation~\cite{Ernst:1974zza}. 
The form factors that take into account the projection of the Pauli-forbidden state to
infinity read,
\begin{eqnarray}
\langle p'|\tilde{t}(E_l)|p_l\rangle \equiv \tilde{h}_{E_l,p_l}(p')  = h_l(p') 
  - \frac{(E_l -E_{p'})}{(E_b-E_{p'})} h_b(p') \frac{1}{E_l -E_b} h_b(p_l) 
  \frac{(E_l -E_{p_l})}{(E_b-E_{p_l})},
\label{eq:b.6}
\end{eqnarray}
where $h_l(p')\equiv t(p',p_l;E_l)$ and $h_b(p')\equiv \langle p'|V|\phi\rangle$. These form factors also define the 
strength constants $\lambda_{nm}$ \cite{Ernst:1974zza}.

\section{Probability of the Pauli Forbidden State in the $^6$Li Ground State Wavefunction}
\label{appendixc}

 The Pauli projection method described in Section~\ref{pauliblocking}
 shifts the energy of the forbidden two-body bound state to positive
 infinity. 
 To estimate the probability of the Pauli-forbidden state in the 
 $^6$Li ground state three-body wavefunction, one needs to project it onto the subspace comprising of product states between the bound
$n\alpha$ pair and the spectator nucleon. The two-body projector 
 $P^{2b}_{\phi}=|\phi_{\beta}^{j_p}\beta\;j_p m_j \rangle 
 \langle \phi_{\beta}^{j_p}\beta\;j_p m_j|$ is defined for that purpose,
where the two-body bound state is characterized by the total
 pair angular momentum $j_p$ and its projection along the $z$-axis $m_{j}$.
 The index $\beta$ represents the
 spins and orbital angular momenta of the pair which couple to $j_p$ separating 
 different angular momentum components of the two-body bound state.  
 Since $P^{2b}_{\phi}$ is defined in the two-body subspace, its application
 in the three-body space requires summation over the spectator quantum numbers
 as well as an integration over the spectator momentum. The projection operator in the three-body space thus takes the form
 \begin{eqnarray}
 P_{\phi}= \sum\limits_{\beta}\sum\limits_{s m_s m_j}\int\limits\;d\vec{q}\;
  \big|\vec{q}\;sm_s\;\phi_{\beta}^{j_p}\;\beta\;j_p m_j \big\rangle\;\big\langle \vec{q}\;sm_s\;\phi_{\beta}^{j_p}\;\beta\;j_p m_j\big|,   
 \label{eq:c1.1}
\end{eqnarray}       
where $s$ is the
  spin of the spectator with $m_s$ being
 its projection along the $z$-axis.  The spectator momentum is denoted
 by $\vec{q}$. Since we are using a partial wave expansion to represent the three-body system, it is advantageous to 
 represent the spectator in terms of partial wave states $|q (\lambda s) {\cal J} m_{\cal J} \rangle$. The angular dependence is 
 expanded in terms of a spectator orbital angular momentum $\lambda$ which is coupled with the spectator spin $s$ to a total 
 spectator angular momentum ${\cal J}$ and its third component  $m_{\cal J}$. Finally, we couple the spectator and two-body 
 bound state angular momenta to the total three-body angular momentum $J$ and its third component $M$. In terms of 
 these states, the projector can  be  rewritten as 
  \begin{eqnarray}
 P_{\phi}= \sum\limits_{\beta}\sum\limits_{\lambda s {\cal J} J M  }\int\limits_{0}^{\infty} \;dq \; q^{2}
  \big|q (j_{p} (\lambda s) {\cal J} ) J M \;\phi_{\beta}^{j_p}\;\beta\; \big\rangle\;\big\langle q (j_{p} (\lambda s) {\cal J} ) J M  \;\phi_{\beta}^{j_p}\;\beta\;\big|,   
 \label{eq:c1.1b}
\end{eqnarray}   
The sum over the angular momentum quantum numbers $\beta$, $\lambda$, $s$,  ${\cal J}$,  $J$ and $M$ and the integral
over $q$ implies that all possible configurations the spectator and the bound pair are 
included.  
Applying $P_{\phi}$ to the three body wavefunction
 yields
\begin{equation}
|\Psi^{P} \rangle \equiv P_{\phi} |\Psi^{JM} \rangle.
\label{eq:c1.2}
\end{equation}
From Eq.~(\ref{eq:c1.2}) we see that the probability of the state $\big|\phi_{\beta}^{j_p}\;\beta\;j_p m_j\big\rangle$ is given by
\begin{eqnarray}
{\cal P}_{TDB}&=&\|\Psi^{P}\|^2\cr\cr
              &=& \big\langle \Psi^{JM}\big|P_{\phi}^\dagger\;P_{\phi}\big|\Psi^{JM}\big\rangle.   
\label{eq:c1.3}
\end{eqnarray}
Since $P_{\phi}$ is Hermitian and fulfills $P_{\phi}^2=P_{\phi}$,
the probability can be recast as the expectation value
\begin{eqnarray}
{\cal P}_{TDB}&=&\big\langle \Psi^{JM}\big|\;P_{\phi}\big|\Psi^{JM}\big\rangle,\cr\cr
& = & \sum\limits_{\beta}\sum\limits_{\lambda s {\cal J} J M  }\int\limits_{0}^{\infty} \;dq \; q^{2}
  \big\langle \Psi^{JM}\big|q (j_{p} (\lambda s) {\cal J} ) J M \;\phi_{\beta}^{j_p}\;\beta\; \big\rangle\;\big\langle q (j_{p} (\lambda s) {\cal J} ) J M  \;\phi_{\beta}^{j_p}\;\beta\;\big| \Psi^{JM}\big\rangle  
 \label{eq:c1.4}
\end{eqnarray}
To evaluate the quantity $\big\langle q (j_{p} (\lambda s) {\cal J} ) J M  \;\phi_{\beta}^{j_p}\;\beta\;\big| \Psi^{JM}\big\rangle $ we insert a complete set of momentum eigenstates
\begin{eqnarray}
\big\langle q (j_{p} (\lambda s) {\cal J} ) J M  \;\phi_{\beta}^{j_p}\;\beta\;\big| \Psi^{JM}\big\rangle &=&
\sum\limits_{\alpha'}\int\limits_{0}^\infty\;dp' {p'}^2\;dq' {q'}^2
\big\langle q (j_{p} (\lambda s) {\cal J} ) J M  \;\phi_{\beta}^{j_p}\;\beta\;\big|
p' q' \alpha' JM \big\rangle\big\langle p' q' \alpha' JM\big|\Psi^{JM}\big\rangle,\cr\cr\cr
&=& \int\limits_{0}^\infty\;dp' {p'}^2\; \phi_{\beta}^{j_p}(p')\; \Psi_{\alpha}^{J}(p'q'). 
\label{eq:c1.7}
\end{eqnarray}
In the last step, we used that the quantum numbers of the projector agree with the definition of 
$\alpha'$ and therefore uniquely define all quantum numbers. The wave functions are independent 
of the third component and the magnitude of the spectator momentum is fixed by the projector, too. 
Based on this result, the desired probability is 
\begin{eqnarray}
{\cal P}_{TDB}&=&\big\langle \Psi^{JM}\big|\;P_{\phi}\big|\Psi^{JM}\big\rangle,\cr\cr
& = & \sum\limits_{\alpha}\int\limits_{0}^{\infty} \;dq \; q^{2}
\left| \int\limits_{0}^\infty\;dp' {p'}^2\; \phi_{\beta}^{j_p}(p')\; \Psi_{\alpha}^{J}(p'q') \right|^{2}
 \label{eq:c1.8}
\end{eqnarray}
where $\alpha$ is short hand notation for $\beta$ and $\lambda$, $s$, ${\cal J}$, $J$ and  $M$ as defined in 
Eq.~(\ref{eq:1.7}) in agreement with Eq.~(\ref{eq:PFstate}).


\begin{acknowledgments}
This work was performed in part under the
auspices of the National Science Foundation under contract NSF-PHY-1520972
with Ohio University and NSF-PHY-1520929 with Michigan State University, 
of the U.~S.  Department of Energy under contract
No. DE-FG02-93ER40756 with Ohio University, and of DFG and NSFC through funds provided to the
Sino-German CRC 110 ``Symmetries and the Emergence of Structure in QCD" (NSFC
Grant No.~11621131001, DFG Grant No.~TRR110).
A.N. acknowledges support of Ohio University through
the Robert and Ren{\'e} Glidden Visiting Professorship Program, and 
 the Institute of Nuclear and Particle
Physics and the Department of Physics and Astronomy.
The authors thank J. Haidenbauer for constructive comments on the manuscript.
 Part of the numerical computations have been 
performed on JUQUEEN and JURECA of the JSC, J\"ulich, Germany. This research also
used resources of the National Energy Research Scientific Computing Center, a DOE
Office of Science User Facility supported by the Office of Science of the U.S. Department of Energy under contract No. DE-AC02-05CH11231.

\end{acknowledgments}


\bibliography{reactions}

\begin{thebibliography}{49}%
\makeatletter
\providecommand \@ifxundefined [1]{%
 \@ifx{#1\undefined}
}%
\providecommand \@ifnum [1]{%
 \ifnum #1\expandafter \@firstoftwo
 \else \expandafter \@secondoftwo
 \fi
}%
\providecommand \@ifx [1]{%
 \ifx #1\expandafter \@firstoftwo
 \else \expandafter \@secondoftwo
 \fi
}%
\providecommand \natexlab [1]{#1}%
\providecommand \enquote  [1]{``#1''}%
\providecommand \bibnamefont  [1]{#1}%
\providecommand \bibfnamefont [1]{#1}%
\providecommand \citenamefont [1]{#1}%
\providecommand \href@noop [0]{\@secondoftwo}%
\providecommand \href [0]{\begingroup \@sanitize@url \@href}%
\providecommand \@href[1]{\@@startlink{#1}\@@href}%
\providecommand \@@href[1]{\endgroup#1\@@endlink}%
\providecommand \@sanitize@url [0]{\catcode `\\12\catcode `\$12\catcode
  `\&12\catcode `\#12\catcode `\^12\catcode `\_12\catcode `\%12\relax}%
\providecommand \@@startlink[1]{}%
\providecommand \@@endlink[0]{}%
\providecommand \url  [0]{\begingroup\@sanitize@url \@url }%
\providecommand \@url [1]{\endgroup\@href {#1}{\urlprefix }}%
\providecommand \urlprefix  [0]{URL }%
\providecommand \Eprint [0]{\href }%
\providecommand \doibase [0]{http://dx.doi.org/}%
\providecommand \selectlanguage [0]{\@gobble}%
\providecommand \bibinfo  [0]{\@secondoftwo}%
\providecommand \bibfield  [0]{\@secondoftwo}%
\providecommand \translation [1]{[#1]}%
\providecommand \BibitemOpen [0]{}%
\providecommand \bibitemStop [0]{}%
\providecommand \bibitemNoStop [0]{.\EOS\space}%
\providecommand \EOS [0]{\spacefactor3000\relax}%
\providecommand \BibitemShut  [1]{\csname bibitem#1\endcsname}%
\let\auto@bib@innerbib\@empty
\bibitem [{\citenamefont {Cizewski}\ \emph {et~al.}(2013)\citenamefont
  {Cizewski} \emph {et~al.}}]{jolie}%
  \BibitemOpen
  \bibfield  {author} {\bibinfo {author} {\bibfnamefont {J.}~\bibnamefont
  {Cizewski}} \emph {et~al.},\ }\href@noop {} {\bibfield  {journal} {\bibinfo
  {journal} {J. Phys. Conf}\ }\textbf {\bibinfo {volume} {420}},\ \bibinfo
  {pages} {012058} (\bibinfo {year} {2013})}\BibitemShut {NoStop}%
\bibitem [{\citenamefont {Kozub}\ \emph {et~al.}(2012)\citenamefont {Kozub},
  \citenamefont {Arbanas}, \citenamefont {Adekola}, \citenamefont {Bardayan},
  \citenamefont {Blackmon} \emph {et~al.}}]{Kozub:2012ka}%
  \BibitemOpen
  \bibfield  {author} {\bibinfo {author} {\bibfnamefont {R.}~\bibnamefont
  {Kozub}}, \bibinfo {author} {\bibfnamefont {G.}~\bibnamefont {Arbanas}},
  \bibinfo {author} {\bibfnamefont {A.}~\bibnamefont {Adekola}}, \bibinfo
  {author} {\bibfnamefont {D.}~\bibnamefont {Bardayan}}, \bibinfo {author}
  {\bibfnamefont {J.}~\bibnamefont {Blackmon}},  \emph {et~al.},\ }\href
  {\doibase 10.1103/PhysRevLett.109.172501} {\bibfield  {journal} {\bibinfo
  {journal} {Phys.Rev.Lett.}\ }\textbf {\bibinfo {volume} {109}},\ \bibinfo
  {pages} {172501} (\bibinfo {year} {2012})}\BibitemShut {NoStop}%
\bibitem [{\citenamefont {Escher}\ \emph {et~al.}(2012)\citenamefont {Escher},
  \citenamefont {Burke}, \citenamefont {Dietrich}, \citenamefont {Scielzo},
  \citenamefont {Thompson},\ and\ \citenamefont {Younes}}]{RevModPhys.84.353}%
  \BibitemOpen
  \bibfield  {author} {\bibinfo {author} {\bibfnamefont {J.~E.}\ \bibnamefont
  {Escher}}, \bibinfo {author} {\bibfnamefont {J.~T.}\ \bibnamefont {Burke}},
  \bibinfo {author} {\bibfnamefont {F.~S.}\ \bibnamefont {Dietrich}}, \bibinfo
  {author} {\bibfnamefont {N.~D.}\ \bibnamefont {Scielzo}}, \bibinfo {author}
  {\bibfnamefont {I.~J.}\ \bibnamefont {Thompson}}, \ and\ \bibinfo {author}
  {\bibfnamefont {W.}~\bibnamefont {Younes}},\ }\href {\doibase
  10.1103/RevModPhys.84.353} {\bibfield  {journal} {\bibinfo  {journal} {Rev.
  Mod. Phys.}\ }\textbf {\bibinfo {volume} {84}},\ \bibinfo {pages} {353}
  (\bibinfo {year} {2012})}\BibitemShut {NoStop}%
\bibitem [{\citenamefont {Potel~{\it et al.}}(2017)}]{Potel:2017z}%
  \BibitemOpen
  \bibfield  {author} {\bibinfo {author} {\bibfnamefont {G.}~\bibnamefont
  {Potel~{\it et al.}}},\ }\href {\doibase 10.1140/epja/i2017-12371-9}
  {\bibfield  {journal} {\bibinfo  {journal} {Eur. Phys. J.}\ }\textbf
  {\bibinfo {volume} {A53}},\ \bibinfo {pages} {178} (\bibinfo {year}
  {2017})}\BibitemShut {NoStop}%
\bibitem [{\citenamefont {Pain}\ \emph {et~al.}(2015)\citenamefont {Pain},
  \citenamefont {Bardayan}, \citenamefont {Blackmon}, \citenamefont {Brown},
  \citenamefont {Chae}, \citenamefont {Chipps}, \citenamefont {Cizewski},
  \citenamefont {Jones}, \citenamefont {Kozub}, \citenamefont {Liang},
  \citenamefont {Matei}, \citenamefont {Matos}, \citenamefont {Moazen},
  \citenamefont {Nesaraja}, \citenamefont {Oko\l{}owicz}, \citenamefont
  {O'Malley}, \citenamefont {Peters}, \citenamefont {Pittman}, \citenamefont
  {P\l{}oszajczak}, \citenamefont {Schmitt}, \citenamefont {Shriner},
  \citenamefont {Shapira}, \citenamefont {Smith}, \citenamefont {Stracener},\
  and\ \citenamefont {Wilson}}]{Pain:2015}%
  \BibitemOpen
  \bibfield  {author} {\bibinfo {author} {\bibfnamefont {S.~D.}\ \bibnamefont
  {Pain}}, \bibinfo {author} {\bibfnamefont {D.~W.}\ \bibnamefont {Bardayan}},
  \bibinfo {author} {\bibfnamefont {J.~C.}\ \bibnamefont {Blackmon}}, \bibinfo
  {author} {\bibfnamefont {S.~M.}\ \bibnamefont {Brown}}, \bibinfo {author}
  {\bibfnamefont {K.~Y.}\ \bibnamefont {Chae}}, \bibinfo {author}
  {\bibfnamefont {K.~A.}\ \bibnamefont {Chipps}}, \bibinfo {author}
  {\bibfnamefont {J.~A.}\ \bibnamefont {Cizewski}}, \bibinfo {author}
  {\bibfnamefont {K.~L.}\ \bibnamefont {Jones}}, \bibinfo {author}
  {\bibfnamefont {R.~L.}\ \bibnamefont {Kozub}}, \bibinfo {author}
  {\bibfnamefont {J.~F.}\ \bibnamefont {Liang}}, \bibinfo {author}
  {\bibfnamefont {C.}~\bibnamefont {Matei}}, \bibinfo {author} {\bibfnamefont
  {M.}~\bibnamefont {Matos}}, \bibinfo {author} {\bibfnamefont {B.~H.}\
  \bibnamefont {Moazen}}, \bibinfo {author} {\bibfnamefont {C.~D.}\
  \bibnamefont {Nesaraja}}, \bibinfo {author} {\bibfnamefont {J.}~\bibnamefont
  {Oko\l{}owicz}}, \bibinfo {author} {\bibfnamefont {P.~D.}\ \bibnamefont
  {O'Malley}}, \bibinfo {author} {\bibfnamefont {W.~A.}\ \bibnamefont
  {Peters}}, \bibinfo {author} {\bibfnamefont {S.~T.}\ \bibnamefont {Pittman}},
  \bibinfo {author} {\bibfnamefont {M.}~\bibnamefont {P\l{}oszajczak}},
  \bibinfo {author} {\bibfnamefont {K.~T.}\ \bibnamefont {Schmitt}}, \bibinfo
  {author} {\bibfnamefont {J.~F.}\ \bibnamefont {Shriner}}, \bibinfo {author}
  {\bibfnamefont {D.}~\bibnamefont {Shapira}}, \bibinfo {author} {\bibfnamefont
  {M.~S.}\ \bibnamefont {Smith}}, \bibinfo {author} {\bibfnamefont {D.~W.}\
  \bibnamefont {Stracener}}, \ and\ \bibinfo {author} {\bibfnamefont {G.~L.}\
  \bibnamefont {Wilson}},\ }\href {\doibase 10.1103/PhysRevLett.114.212501}
  {\bibfield  {journal} {\bibinfo  {journal} {Phys. Rev. Lett.}\ }\textbf
  {\bibinfo {volume} {114}},\ \bibinfo {pages} {212501} (\bibinfo {year}
  {2015})}\BibitemShut {NoStop}%
\bibitem [{\citenamefont {Schmitt}\ \emph {et~al.}(2012)\citenamefont
  {Schmitt}, \citenamefont {Jones}, \citenamefont {Bey}, \citenamefont {Ahn},
  \citenamefont {Bardayan} \emph {et~al.}}]{Schmitt:2012bt}%
  \BibitemOpen
  \bibfield  {author} {\bibinfo {author} {\bibfnamefont {K.~T.}\ \bibnamefont
  {Schmitt}}, \bibinfo {author} {\bibfnamefont {K.~L.}\ \bibnamefont {Jones}},
  \bibinfo {author} {\bibfnamefont {A.}~\bibnamefont {Bey}}, \bibinfo {author}
  {\bibfnamefont {S.}~\bibnamefont {Ahn}}, \bibinfo {author} {\bibfnamefont
  {D.}~\bibnamefont {Bardayan}},  \emph {et~al.},\ }\href {\doibase
  10.1103/PhysRevLett.108.192701} {\bibfield  {journal} {\bibinfo  {journal}
  {Phys.Rev.Lett.}\ }\textbf {\bibinfo {volume} {108}},\ \bibinfo {pages}
  {192701} (\bibinfo {year} {2012})}\BibitemShut {NoStop}%
\bibitem [{\citenamefont {Jones}\ \emph {et~al.}(2011)\citenamefont {Jones},
  \citenamefont {Nunes}, \citenamefont {Adekola}, \citenamefont {Bardayan},
  \citenamefont {Blackmon} \emph {et~al.}}]{Jones:2011kp}%
  \BibitemOpen
  \bibfield  {author} {\bibinfo {author} {\bibfnamefont {K.~L.}\ \bibnamefont
  {Jones}}, \bibinfo {author} {\bibfnamefont {F.~M.}\ \bibnamefont {Nunes}},
  \bibinfo {author} {\bibfnamefont {A.}~\bibnamefont {Adekola}}, \bibinfo
  {author} {\bibfnamefont {D.}~\bibnamefont {Bardayan}}, \bibinfo {author}
  {\bibfnamefont {J.}~\bibnamefont {Blackmon}},  \emph {et~al.},\ }\href
  {\doibase 10.1103/PhysRevC.84.034601} {\bibfield  {journal} {\bibinfo
  {journal} {Phys.Rev.}\ }\textbf {\bibinfo {volume} {C84}},\ \bibinfo {pages}
  {034601} (\bibinfo {year} {2011})}\BibitemShut {NoStop}%
\bibitem [{\citenamefont {Deltuva}(2013)}]{Deltuva2013}%
  \BibitemOpen
  \bibfield  {author} {\bibinfo {author} {\bibfnamefont {A.}~\bibnamefont
  {Deltuva}},\ }\href {\doibase 10.1103/PhysRevC.88.011601} {\bibfield
  {journal} {\bibinfo  {journal} {Phys. Rev. C}\ }\textbf {\bibinfo {volume}
  {88}},\ \bibinfo {pages} {011601} (\bibinfo {year} {2013})}\BibitemShut
  {NoStop}%
\bibitem [{\citenamefont {Varner}\ \emph {et~al.}(1991)\citenamefont {Varner},
  \citenamefont {Thompson}, \citenamefont {McAbee}, \citenamefont {Ludwig},\
  and\ \citenamefont {Clegg}}]{Varner:1991zz}%
  \BibitemOpen
  \bibfield  {author} {\bibinfo {author} {\bibfnamefont {R.}~\bibnamefont
  {Varner}}, \bibinfo {author} {\bibfnamefont {W.}~\bibnamefont {Thompson}},
  \bibinfo {author} {\bibfnamefont {T.}~\bibnamefont {McAbee}}, \bibinfo
  {author} {\bibfnamefont {E.}~\bibnamefont {Ludwig}}, \ and\ \bibinfo {author}
  {\bibfnamefont {T.}~\bibnamefont {Clegg}},\ }\href {\doibase 10.10
  16/0370-1573(91)90039-O} {\bibfield  {journal} {\bibinfo  {journal}
  {Phys.Rept.}\ }\textbf {\bibinfo {volume} {201}},\ \bibinfo {pages} {57}
  (\bibinfo {year} {1991})}\BibitemShut {NoStop}%
\bibitem [{\citenamefont {Weppner}\ \emph {et~al.}(2009)\citenamefont
  {Weppner}, \citenamefont {Penney}, \citenamefont {Diffendale},\ and\
  \citenamefont {Vittorini}}]{Weppner:2009qy}%
  \BibitemOpen
  \bibfield  {author} {\bibinfo {author} {\bibfnamefont {S.}~\bibnamefont
  {Weppner}}, \bibinfo {author} {\bibfnamefont {R.}~\bibnamefont {Penney}},
  \bibinfo {author} {\bibfnamefont {G.}~\bibnamefont {Diffendale}}, \ and\
  \bibinfo {author} {\bibfnamefont {G.}~\bibnamefont {Vittorini}},\ }\href
  {\doibase 10.1103/PhysRevC.80.034608} {\bibfield  {journal} {\bibinfo
  {journal} {Phys.Rev.}\ }\textbf {\bibinfo {volume} {C80}},\ \bibinfo {pages}
  {034608} (\bibinfo {year} {2009})}\BibitemShut {NoStop}%
\bibitem [{\citenamefont {Koning}\ and\ \citenamefont
  {Delaroche}(2003)}]{Koning:2003zz}%
  \BibitemOpen
  \bibfield  {author} {\bibinfo {author} {\bibfnamefont {A.}~\bibnamefont
  {Koning}}\ and\ \bibinfo {author} {\bibfnamefont {J.}~\bibnamefont
  {Delaroche}},\ }\href@noop {} {\bibfield  {journal} {\bibinfo  {journal}
  {Nucl.Phys.}\ }\textbf {\bibinfo {volume} {A713}},\ \bibinfo {pages} {231}
  (\bibinfo {year} {2003})}\BibitemShut {NoStop}%
\bibitem [{\citenamefont {Deltuva}\ and\ \citenamefont
  {Fonseca}(2009)}]{Deltuva:2009fp}%
  \BibitemOpen
  \bibfield  {author} {\bibinfo {author} {\bibfnamefont {A.}~\bibnamefont
  {Deltuva}}\ and\ \bibinfo {author} {\bibfnamefont {A.}~\bibnamefont
  {Fonseca}},\ }\href {\doibase 10.1103/PhysRevC.79.014606} {\bibfield
  {journal} {\bibinfo  {journal} {Phys.Rev.}\ }\textbf {\bibinfo {volume}
  {C79}},\ \bibinfo {pages} {014606} (\bibinfo {year} {2009})}\BibitemShut
  {NoStop}%
\bibitem [{\citenamefont {Deltuva}(2009)}]{Deltuva:2009cr}%
  \BibitemOpen
  \bibfield  {author} {\bibinfo {author} {\bibfnamefont {A.}~\bibnamefont
  {Deltuva}},\ }\href {\doibase 10.1103/PhysRevC.79.054603} {\bibfield
  {journal} {\bibinfo  {journal} {Phys.Rev.}\ }\textbf {\bibinfo {volume}
  {C79}},\ \bibinfo {pages} {054603} (\bibinfo {year} {2009})}\BibitemShut
  {NoStop}%
\bibitem [{\citenamefont {Nunes}\ and\ \citenamefont
  {Deltuva}(2011)}]{Nunes:2011cv}%
  \BibitemOpen
  \bibfield  {author} {\bibinfo {author} {\bibfnamefont {F.~M.}\ \bibnamefont
  {Nunes}}\ and\ \bibinfo {author} {\bibfnamefont {A.}~\bibnamefont
  {Deltuva}},\ }\href {\doibase 10.1103/PhysRevC.84.034607} {\bibfield
  {journal} {\bibinfo  {journal} {Phys.Rev.}\ }\textbf {\bibinfo {volume}
  {C84}},\ \bibinfo {pages} {034607} (\bibinfo {year} {2011})}\BibitemShut
  {NoStop}%
\bibitem [{\citenamefont {Upadhyay}\ \emph {et~al.}(2012)\citenamefont
  {Upadhyay}, \citenamefont {Deltuva},\ and\ \citenamefont
  {Nunes}}]{Upadhyay:2011ta}%
  \BibitemOpen
  \bibfield  {author} {\bibinfo {author} {\bibfnamefont {N.~J.}\ \bibnamefont
  {Upadhyay}}, \bibinfo {author} {\bibfnamefont {A.}~\bibnamefont {Deltuva}}, \
  and\ \bibinfo {author} {\bibfnamefont {F.~M.}\ \bibnamefont {Nunes}},\ }\href
  {\doibase 10.1103/PhysRevC.85.054621} {\bibfield  {journal} {\bibinfo
  {journal} {Phys.Rev.}\ }\textbf {\bibinfo {volume} {C85}},\ \bibinfo {pages}
  {054621} (\bibinfo {year} {2012})}, \BibitemShut {NoStop}%
\bibitem [{\citenamefont {Mukhamedzhanov}\ \emph {et~al.}(2012)\citenamefont
  {Mukhamedzhanov}, \citenamefont {Eremenko},\ and\ \citenamefont
  {Sattarov}}]{Mukhamedzhanov:2012qv}%
  \BibitemOpen
  \bibfield  {author} {\bibinfo {author} {\bibfnamefont {A.}~\bibnamefont
  {Mukhamedzhanov}}, \bibinfo {author} {\bibfnamefont {V.}~\bibnamefont
  {Eremenko}}, \ and\ \bibinfo {author} {\bibfnamefont {A.}~\bibnamefont
  {Sattarov}},\ }\href {\doibase 10.1103/PhysRevC.86.034001} {\bibfield
  {journal} {\bibinfo  {journal} {Phys.Rev.}\ }\textbf {\bibinfo {volume}
  {C86}},\ \bibinfo {pages} {034001} (\bibinfo {year} {2012})}\BibitemShut
  {NoStop}%
\bibitem [{\citenamefont {Eskandarian}\ and\ \citenamefont
  {Afnan}(1992)}]{Eskandarian:1992zz}%
  \BibitemOpen
  \bibfield  {author} {\bibinfo {author} {\bibfnamefont {A.}~\bibnamefont
  {Eskandarian}}\ and\ \bibinfo {author} {\bibfnamefont {I.~R.}\ \bibnamefont
  {Afnan}},\ }\href {\doibase 10.1103/PhysRevC.46.2344} {\bibfield  {journal}
  {\bibinfo  {journal} {Phys. Rev.}\ }\textbf {\bibinfo {volume} {C46}},\
  \bibinfo {pages} {2344} (\bibinfo {year} {1992})}\BibitemShut {NoStop}%
\bibitem [{\citenamefont {Lehman}(1982)}]{Lehman:1982zz}%
  \BibitemOpen
  \bibfield  {author} {\bibinfo {author} {\bibfnamefont {D.~R.}\ \bibnamefont
  {Lehman}},\ }\href {\doibase 10.1103/PhysRevC.25.3146} {\bibfield  {journal}
  {\bibinfo  {journal} {Phys. Rev.}\ }\textbf {\bibinfo {volume} {C25}},\
  \bibinfo {pages} {3146} (\bibinfo {year} {1982})}\BibitemShut {NoStop}%
\bibitem [{\citenamefont {Lehman}\ \emph {et~al.}(1978)\citenamefont {Lehman},
  \citenamefont {Rai},\ and\ \citenamefont {Ghovanlou}}]{Lehman:1978zz}%
  \BibitemOpen
  \bibfield  {author} {\bibinfo {author} {\bibfnamefont {D.~R.}\ \bibnamefont
  {Lehman}}, \bibinfo {author} {\bibfnamefont {M.}~\bibnamefont {Rai}}, \ and\
  \bibinfo {author} {\bibfnamefont {A.}~\bibnamefont {Ghovanlou}},\ }\href
  {\doibase 10.1103/PhysRevC.17.744} {\bibfield  {journal} {\bibinfo  {journal}
  {Phys. Rev.}\ }\textbf {\bibinfo {volume} {C17}},\ \bibinfo {pages} {744}
  (\bibinfo {year} {1978})}\BibitemShut {NoStop}%
\bibitem [{\citenamefont {Ghovanlou}\ and\ \citenamefont
  {Lehman}(1974)}]{Ghovanlou:1974zza}%
  \BibitemOpen
  \bibfield  {author} {\bibinfo {author} {\bibfnamefont {A.}~\bibnamefont
  {Ghovanlou}}\ and\ \bibinfo {author} {\bibfnamefont {D.~R.}\ \bibnamefont
  {Lehman}},\ }\href {\doibase 10.1103/PhysRevC.9.1730} {\bibfield  {journal}
  {\bibinfo  {journal} {Phys. Rev.}\ }\textbf {\bibinfo {volume} {C9}},\
  \bibinfo {pages} {1730} (\bibinfo {year} {1974})}\BibitemShut {NoStop}%
\bibitem [{\citenamefont {Haidenbauer}\ and\ \citenamefont
  {Plessas}(1983)}]{Haidenbauer:1982if}%
  \BibitemOpen
  \bibfield  {author} {\bibinfo {author} {\bibfnamefont {J.}~\bibnamefont
  {Haidenbauer}}\ and\ \bibinfo {author} {\bibfnamefont {W.}~\bibnamefont
  {Plessas}},\ }\href {\doibase 10.1103/PhysRevC.27.63} {\bibfield  {journal}
  {\bibinfo  {journal} {Phys.Rev.}\ }\textbf {\bibinfo {volume} {C27}},\
  \bibinfo {pages} {63} (\bibinfo {year} {1983})}\BibitemShut {NoStop}%
\bibitem [{\citenamefont {Haidenbauer}\ \emph {et~al.}(1986)\citenamefont
  {Haidenbauer}, \citenamefont {Koike},\ and\ \citenamefont
  {Plessas}}]{Haidenbauer:1986zza}%
  \BibitemOpen
  \bibfield  {author} {\bibinfo {author} {\bibfnamefont {J.}~\bibnamefont
  {Haidenbauer}}, \bibinfo {author} {\bibfnamefont {Y.}~\bibnamefont {Koike}},
  \ and\ \bibinfo {author} {\bibfnamefont {W.}~\bibnamefont {Plessas}},\ }\href
  {\doibase 10.1103/PhysRevC.33.439} {\bibfield  {journal} {\bibinfo  {journal}
  {Phys.Rev.}\ }\textbf {\bibinfo {volume} {C33}},\ \bibinfo {pages} {439}
  (\bibinfo {year} {1986})}\BibitemShut {NoStop}%
\bibitem [{\citenamefont {Ernst}\ \emph {et~al.}(1973)\citenamefont {Ernst},
  \citenamefont {Shakin},\ and\ \citenamefont {Thaler}}]{Ernst:1973zzb}%
  \BibitemOpen
  \bibfield  {author} {\bibinfo {author} {\bibfnamefont {D.~J.}\ \bibnamefont
  {Ernst}}, \bibinfo {author} {\bibfnamefont {C.~M.}\ \bibnamefont {Shakin}}, \
  and\ \bibinfo {author} {\bibfnamefont {R.~M.}\ \bibnamefont {Thaler}},\
  }\href {\doibase 10.1103/PhysRevC.8.46} {\bibfield  {journal} {\bibinfo
  {journal} {Phys.Rev.}\ }\textbf {\bibinfo {volume} {C8}},\ \bibinfo {pages}
  {46} (\bibinfo {year} {1973})}\BibitemShut {NoStop}%
\bibitem [{\citenamefont {Hlophe}\ \emph {et~al.}(2013)\citenamefont {Hlophe}
  \emph {et~al.}}]{Hlophe:2013xca}%
  \BibitemOpen
  \bibfield  {author} {\bibinfo {author} {\bibfnamefont {L.}~\bibnamefont
  {Hlophe}} \emph {et~al.} (\bibinfo {collaboration} {The TORUS
  Collaboration}),\ }\href {\doibase 10.1103/PhysRevC.88.064608} {\bibfield
  {journal} {\bibinfo  {journal} {Phys.Rev.}\ }\textbf {\bibinfo {volume}
  {C88}},\ \bibinfo {pages} {064608} (\bibinfo {year} {2013})} \BibitemShut
  {NoStop}%
\bibitem [{\citenamefont {Hlophe}\ \emph {et~al.}(2014)\citenamefont {Hlophe}
  \emph {et~al.}}]{Hlophe:2014xca}%
  \BibitemOpen
  \bibfield  {author} {\bibinfo {author} {\bibfnamefont {L.}~\bibnamefont
  {Hlophe}} \emph {et~al.} (\bibinfo {collaboration} {The TORUS
  Collaboration}),\ }\href {\doibase 10.1103/PhysRevC.90.061602} {\bibfield
  {journal} {\bibinfo  {journal} {Phys.Rev.}\ }\textbf {\bibinfo {volume}
  {C90}},\ \bibinfo {pages} {061602(R)} (\bibinfo {year} {2014})} \BibitemShut
  {NoStop}%
\bibitem [{\citenamefont {Hlophe}\ and\ \citenamefont
  {Elster}(2016)}]{Hlophe:2015rqn}%
  \BibitemOpen
  \bibfield  {author} {\bibinfo {author} {\bibfnamefont {L.}~\bibnamefont
  {Hlophe}}\ and\ \bibinfo {author} {\bibfnamefont {C.}~\bibnamefont
  {Elster}},\ }\href {\doibase 10.1103/PhysRevC.93.034601} {\bibfield
  {journal} {\bibinfo  {journal} {Phys. Rev.}\ }\textbf {\bibinfo {volume}
  {C93}},\ \bibinfo {pages} {034601} (\bibinfo {year} {2016})} \BibitemShut
  {NoStop}%
\bibitem [{\citenamefont {Cornelius}\ \emph {et~al.}(1990)\citenamefont
  {Cornelius}, \citenamefont {Gloeckle}, \citenamefont {Haidenbauer},
  \citenamefont {Koike}, \citenamefont {Plessas},\ and\ \citenamefont
  {Witala}}]{Cornelius:1990zz}%
  \BibitemOpen
  \bibfield  {author} {\bibinfo {author} {\bibfnamefont {T.}~\bibnamefont
  {Cornelius}}, \bibinfo {author} {\bibfnamefont {W.}~\bibnamefont {Gloeckle}},
  \bibinfo {author} {\bibfnamefont {J.}~\bibnamefont {Haidenbauer}}, \bibinfo
  {author} {\bibfnamefont {Y.}~\bibnamefont {Koike}}, \bibinfo {author}
  {\bibfnamefont {W.}~\bibnamefont {Plessas}}, \ and\ \bibinfo {author}
  {\bibfnamefont {H.}~\bibnamefont {Witala}},\ }\href {\doibase
  10.1103/PhysRevC.41.2538} {\bibfield  {journal} {\bibinfo  {journal} {Phys.
  Rev.}\ }\textbf {\bibinfo {volume} {C41}},\ \bibinfo {pages} {2538} (\bibinfo
  {year} {1990})}\BibitemShut {NoStop}%
\bibitem [{\citenamefont {Nemoto}\ \emph {et~al.}(1998)\citenamefont {Nemoto},
  \citenamefont {Chmielewski}, \citenamefont {Schellingerhout}, \citenamefont
  {Sauer}, \citenamefont {Haidenbauer},\ and\ \citenamefont
  {Oryu}}]{Nemoto:1998wt}%
  \BibitemOpen
  \bibfield  {author} {\bibinfo {author} {\bibfnamefont {S.}~\bibnamefont
  {Nemoto}}, \bibinfo {author} {\bibfnamefont {K.}~\bibnamefont {Chmielewski}},
  \bibinfo {author} {\bibfnamefont {N.~W.}\ \bibnamefont {Schellingerhout}},
  \bibinfo {author} {\bibfnamefont {P.~U.}\ \bibnamefont {Sauer}}, \bibinfo
  {author} {\bibfnamefont {J.}~\bibnamefont {Haidenbauer}}, \ and\ \bibinfo
  {author} {\bibfnamefont {S.}~\bibnamefont {Oryu}},\ }\href {\doibase
  10.1007/s006010050087} {\bibfield  {journal} {\bibinfo  {journal} {Few Body
  Syst.}\ }\textbf {\bibinfo {volume} {24}},\ \bibinfo {pages} {213} (\bibinfo
  {year} {1998})}\BibitemShut {NoStop}%
\bibitem [{\citenamefont {Kamada}\ \emph {et~al.}(2001)\citenamefont {Kamada},
  \citenamefont {Nogga}, \citenamefont {Gl\"ockle}, \citenamefont {Hiyama},
  \citenamefont {Kamimura}, \citenamefont {Varga}, \citenamefont {Suzuki},
  \citenamefont {Viviani}, \citenamefont {Kievsky}, \citenamefont {Rosati},
  \citenamefont {Carlson}, \citenamefont {Pieper}, \citenamefont {Wiringa},
  \citenamefont {Navr\'atil}, \citenamefont {Barrett}, \citenamefont {Barnea},
  \citenamefont {Leidemann},\ and\ \citenamefont {Orlandini}}]{benchmark1}%
  \BibitemOpen
  \bibfield  {author} {\bibinfo {author} {\bibfnamefont {H.}~\bibnamefont
  {Kamada}}, \bibinfo {author} {\bibfnamefont {A.}~\bibnamefont {Nogga}},
  \bibinfo {author} {\bibfnamefont {W.}~\bibnamefont {Gl\"ockle}}, \bibinfo
  {author} {\bibfnamefont {E.}~\bibnamefont {Hiyama}}, \bibinfo {author}
  {\bibfnamefont {M.}~\bibnamefont {Kamimura}}, \bibinfo {author}
  {\bibfnamefont {K.}~\bibnamefont {Varga}}, \bibinfo {author} {\bibfnamefont
  {Y.}~\bibnamefont {Suzuki}}, \bibinfo {author} {\bibfnamefont
  {M.}~\bibnamefont {Viviani}}, \bibinfo {author} {\bibfnamefont
  {A.}~\bibnamefont {Kievsky}}, \bibinfo {author} {\bibfnamefont
  {S.}~\bibnamefont {Rosati}}, \bibinfo {author} {\bibfnamefont
  {J.}~\bibnamefont {Carlson}}, \bibinfo {author} {\bibfnamefont {S.~C.}\
  \bibnamefont {Pieper}}, \bibinfo {author} {\bibfnamefont {R.~B.}\
  \bibnamefont {Wiringa}}, \bibinfo {author} {\bibfnamefont {P.}~\bibnamefont
  {Navr\'atil}}, \bibinfo {author} {\bibfnamefont {B.~R.}\ \bibnamefont
  {Barrett}}, \bibinfo {author} {\bibfnamefont {N.}~\bibnamefont {Barnea}},
  \bibinfo {author} {\bibfnamefont {W.}~\bibnamefont {Leidemann}}, \ and\
  \bibinfo {author} {\bibfnamefont {G.}~\bibnamefont {Orlandini}},\ }\href
  {\doibase 10.1103/PhysRevC.64.044001} {\bibfield  {journal} {\bibinfo
  {journal} {Phys. Rev. C}\ }\textbf {\bibinfo {volume} {64}},\ \bibinfo
  {pages} {044001} (\bibinfo {year} {2001})}\BibitemShut {NoStop}%
\bibitem [{\citenamefont {Viviani}\ \emph {et~al.}(2017)\citenamefont
  {Viviani}, \citenamefont {Deltuva}, \citenamefont {Lazauskas}, \citenamefont
  {Fonseca}, \citenamefont {Kievsky},\ and\ \citenamefont
  {Marcucci}}]{benchmark2}%
  \BibitemOpen
  \bibfield  {author} {\bibinfo {author} {\bibfnamefont {M.}~\bibnamefont
  {Viviani}}, \bibinfo {author} {\bibfnamefont {A.}~\bibnamefont {Deltuva}},
  \bibinfo {author} {\bibfnamefont {R.}~\bibnamefont {Lazauskas}}, \bibinfo
  {author} {\bibfnamefont {A.~C.}\ \bibnamefont {Fonseca}}, \bibinfo {author}
  {\bibfnamefont {A.}~\bibnamefont {Kievsky}}, \ and\ \bibinfo {author}
  {\bibfnamefont {L.~E.}\ \bibnamefont {Marcucci}},\ }\href {\doibase
  10.1103/PhysRevC.95.034003} {\bibfield  {journal} {\bibinfo  {journal} {Phys.
  Rev. C}\ }\textbf {\bibinfo {volume} {95}},\ \bibinfo {pages} {034003}
  (\bibinfo {year} {2017})}\BibitemShut {NoStop}%
\bibitem [{\citenamefont {Schellingerhout}\ \emph {et~al.}(1993)\citenamefont
  {Schellingerhout}, \citenamefont {Kok}, \citenamefont {Coon},\ and\
  \citenamefont {Adam}}]{Schellingerhout:1993ku}%
  \BibitemOpen
  \bibfield  {author} {\bibinfo {author} {\bibfnamefont {N.~W.}\ \bibnamefont
  {Schellingerhout}}, \bibinfo {author} {\bibfnamefont {L.~P.}\ \bibnamefont
  {Kok}}, \bibinfo {author} {\bibfnamefont {S.~A.}\ \bibnamefont {Coon}}, \
  and\ \bibinfo {author} {\bibfnamefont {R.~M.}\ \bibnamefont {Adam}},\ }\href
  {\doibase 10.1103/PhysRevC.48.2714, 10.1103/PhysRevC.52.439} {\bibfield
  {journal} {\bibinfo  {journal} {Phys. Rev.}\ }\textbf {\bibinfo {volume}
  {C48}},\ \bibinfo {pages} {2714} (\bibinfo {year} {1993})},\ \bibinfo {note}
  {[Erratum: Phys. Rev.C52,439(1995)]}
  \BibitemShut {NoStop}%
\bibitem [{\citenamefont {Balian}\ and\ \citenamefont
  {Brezin}(1969)}]{Balian:1969sd}%
  \BibitemOpen
  \bibfield  {author} {\bibinfo {author} {\bibfnamefont {R.}~\bibnamefont
  {Balian}}\ and\ \bibinfo {author} {\bibfnamefont {E.}~\bibnamefont
  {Brezin}},\ }\href {\doibase 10.1007/BF02710946} {\bibfield  {journal}
  {\bibinfo  {journal} {Nuovo Cim.}\ }\textbf {\bibinfo {volume} {B61}},\
  \bibinfo {pages} {403} (\bibinfo {year} {1969})}\BibitemShut {NoStop}%
\bibitem [{\citenamefont {Schmid}\ and\ \citenamefont
  {Ziegelmann}(1974)}]{SchmidZiegelmann}%
  \BibitemOpen
  \bibfield  {author} {\bibinfo {author} {\bibfnamefont {E.}~\bibnamefont
  {Schmid}}\ and\ \bibinfo {author} {\bibfnamefont {H.}~\bibnamefont
  {Ziegelmann}},\ }\href@noop {} {\emph {\bibinfo {title} {The Quantum
  Mechanical Three-Body Problem}}},\ Vieweg tracts in pure and applied physics\
  (\bibinfo  {publisher} {Elsevier},\ \bibinfo {year} {1974})\BibitemShut
  {NoStop}%
\bibitem [{\citenamefont {Saad}(2003)}]{Saad:2003}%
  \BibitemOpen
  \bibfield  {author} {\bibinfo {author} {\bibfnamefont {Y.}~\bibnamefont
  {Saad}},\ }\href@noop {} {\emph {\bibinfo {title} {{Iterative Methods for
  Sparse Linear Systems}}}}\ (\bibinfo  {publisher} {Society for Industrial and
  Applied Mathematics, SIAM},\ \bibinfo {year} {2003})\BibitemShut {NoStop}%
\bibitem [{\citenamefont {Kukulin}\ and\ \citenamefont
  {Pomerantsev}(1978)}]{Kukulin:1978he}%
  \BibitemOpen
  \bibfield  {author} {\bibinfo {author} {\bibfnamefont {V.~I.}\ \bibnamefont
  {Kukulin}}\ and\ \bibinfo {author} {\bibfnamefont {V.~N.}\ \bibnamefont
  {Pomerantsev}},\ }\href {\doibase 10.1016/0003-4916(78)90069-6} {\bibfield
  {journal} {\bibinfo  {journal} {Annals Phys.}\ }\textbf {\bibinfo {volume}
  {111}},\ \bibinfo {pages} {330} (\bibinfo {year} {1978})}\BibitemShut
  {NoStop}%
\bibitem [{\citenamefont {Thompson}\ \emph {et~al.}(2000)\citenamefont
  {Thompson}, \citenamefont {Danilin}, \citenamefont {Efros}, \citenamefont
  {Vaagen}, \citenamefont {Bang},\ and\ \citenamefont
  {Zhukov}}]{Thompson:2000ny}%
  \BibitemOpen
  \bibfield  {author} {\bibinfo {author} {\bibfnamefont {I.~J.}\ \bibnamefont
  {Thompson}}, \bibinfo {author} {\bibfnamefont {B.~V.}\ \bibnamefont
  {Danilin}}, \bibinfo {author} {\bibfnamefont {V.~D.}\ \bibnamefont {Efros}},
  \bibinfo {author} {\bibfnamefont {J.~S.}\ \bibnamefont {Vaagen}}, \bibinfo
  {author} {\bibfnamefont {J.~M.}\ \bibnamefont {Bang}}, \ and\ \bibinfo
  {author} {\bibfnamefont {M.~V.}\ \bibnamefont {Zhukov}},\ }\href {\doibase
  10.1103/PhysRevC.61.024318} {\bibfield  {journal} {\bibinfo  {journal} {Phys.
  Rev.}\ }\textbf {\bibinfo {volume} {C61}},\ \bibinfo {pages} {024318}
  (\bibinfo {year} {2000})}\BibitemShut {NoStop}%
\bibitem [{\citenamefont {Bang}\ \emph {et~al.}(1983)\citenamefont {Bang},
  \citenamefont {Benayoun}, \citenamefont {Gignoux},\ and\ \citenamefont
  {Thompson}}]{Bang:1983xpz}%
  \BibitemOpen
  \bibfield  {author} {\bibinfo {author} {\bibfnamefont {J.}~\bibnamefont
  {Bang}}, \bibinfo {author} {\bibfnamefont {J.~J.}\ \bibnamefont {Benayoun}},
  \bibinfo {author} {\bibfnamefont {C.}~\bibnamefont {Gignoux}}, \ and\
  \bibinfo {author} {\bibfnamefont {I.~J.}\ \bibnamefont {Thompson}},\ }\href
  {\doibase 10.1016/0375-9474(83)90327-5} {\bibfield  {journal} {\bibinfo
  {journal} {Nucl. Phys.}\ }\textbf {\bibinfo {volume} {A405}},\ \bibinfo
  {pages} {126} (\bibinfo {year} {1983})}\BibitemShut {NoStop}%
\bibitem [{\citenamefont {Gell-Mann}\ and\ \citenamefont
  {Goldberger}(1953)}]{GellMann:1953zz}%
  \BibitemOpen
  \bibfield  {author} {\bibinfo {author} {\bibfnamefont {M.}~\bibnamefont
  {Gell-Mann}}\ and\ \bibinfo {author} {\bibfnamefont {M.~L.}\ \bibnamefont
  {Goldberger}},\ }\href {\doibase 10.1103/PhysRev.91.398} {\bibfield
  {journal} {\bibinfo  {journal} {Phys. Rev.}\ }\textbf {\bibinfo {volume}
  {91}},\ \bibinfo {pages} {398} (\bibinfo {year} {1953})}\BibitemShut
  {NoStop}%
\bibitem [{\citenamefont {Nogga}\ \emph {et~al.}(2005)\citenamefont {Nogga},
  \citenamefont {Timmermans},\ and\ \citenamefont {van Kolck}}]{Nogga:2005hy}%
  \BibitemOpen
  \bibfield  {author} {\bibinfo {author} {\bibfnamefont {A.}~\bibnamefont
  {Nogga}}, \bibinfo {author} {\bibfnamefont {R.~G.~E.}\ \bibnamefont
  {Timmermans}}, \ and\ \bibinfo {author} {\bibfnamefont {U.}~\bibnamefont {van
  Kolck}},\ }\href {\doibase 10.1103/PhysRevC.72.054006} {\bibfield  {journal}
  {\bibinfo  {journal} {Phys. Rev.}\ }\textbf {\bibinfo {volume} {C72}},\
  \bibinfo {pages} {054006} (\bibinfo {year} {2005})}
  \BibitemShut {NoStop}%
\bibitem [{\citenamefont {Ernst}\ \emph {et~al.}(1974)\citenamefont {Ernst},
  \citenamefont {Shakin},\ and\ \citenamefont {Thaler}}]{Ernst:1974zza}%
  \BibitemOpen
  \bibfield  {author} {\bibinfo {author} {\bibfnamefont {D.~J.}\ \bibnamefont
  {Ernst}}, \bibinfo {author} {\bibfnamefont {C.~M.}\ \bibnamefont {Shakin}}, \
  and\ \bibinfo {author} {\bibfnamefont {R.~M.}\ \bibnamefont {Thaler}},\
  }\href {\doibase 10.1103/PhysRevC.9.1780} {\bibfield  {journal} {\bibinfo
  {journal} {Phys.Rev.}\ }\textbf {\bibinfo {volume} {C9}},\ \bibinfo {pages}
  {1780} (\bibinfo {year} {1974})}\BibitemShut {NoStop}%
\bibitem [{\citenamefont {Machleidt}(2001)}]{Machleidt:2000ge}%
  \BibitemOpen
  \bibfield  {author} {\bibinfo {author} {\bibfnamefont {R.}~\bibnamefont
  {Machleidt}},\ }\href {\doibase 10.1103/PhysRevC.63.024001} {\bibfield
  {journal} {\bibinfo  {journal} {Phys. Rev.}\ }\textbf {\bibinfo {volume}
  {C63}},\ \bibinfo {pages} {024001} (\bibinfo {year} {2001})}
  \BibitemShut {NoStop}%
\bibitem [{\citenamefont {Bang}\ and\ \citenamefont
  {Gignoux}(1979)}]{Bang:1979ihm}%
  \BibitemOpen
  \bibfield  {author} {\bibinfo {author} {\bibfnamefont {J.}~\bibnamefont
  {Bang}}\ and\ \bibinfo {author} {\bibfnamefont {C.}~\bibnamefont {Gignoux}},\
  }\href {\doibase 10.1016/0375-9474(79)90571-2} {\bibfield  {journal}
  {\bibinfo  {journal} {Nucl. Phys.}\ }\textbf {\bibinfo {volume} {A313}},\
  \bibinfo {pages} {119} (\bibinfo {year} {1979})}\BibitemShut {NoStop}%
\bibitem [{\citenamefont {Stammbach}\ and\ \citenamefont
  {Walter}(1972)}]{Stammbach72}%
  \BibitemOpen
  \bibfield  {author} {\bibinfo {author} {\bibfnamefont {T.}~\bibnamefont
  {Stammbach}}\ and\ \bibinfo {author} {\bibfnamefont {R.}~\bibnamefont
  {Walter}},\ }\href {\doibase http://dx.doi.org/10.1016/0375-9474(72)90166-2}
  {\bibfield  {journal} {\bibinfo  {journal} {Nuclear Physics A}\ }\textbf
  {\bibinfo {volume} {180}},\ \bibinfo {pages} {225 } (\bibinfo {year}
  {1972})}\BibitemShut {NoStop}%
\bibitem [{\citenamefont {Ajzenberg-Selove}(1988)}]{AjzenbergSelove:1988ec}%
  \BibitemOpen
  \bibfield  {author} {\bibinfo {author} {\bibfnamefont {F.}~\bibnamefont
  {Ajzenberg-Selove}},\ }\href {\doibase 10.1016/0375-9474(88)90124-8}
  {\bibfield  {journal} {\bibinfo  {journal} {Nucl. Phys.}\ }\textbf {\bibinfo
  {volume} {A490}},\ \bibinfo {pages} {1} (\bibinfo {year} {1988})}\BibitemShut
  {NoStop}%
\bibitem [{\citenamefont {Zhukov}\ \emph {et~al.}(1993)\citenamefont {Zhukov},
  \citenamefont {Danilin}, \citenamefont {Fedorov}, \citenamefont {Bang},
  \citenamefont {Thompson},\ and\ \citenamefont {Vaagen}}]{Zhukov:1993aw}%
  \BibitemOpen
  \bibfield  {author} {\bibinfo {author} {\bibfnamefont {M.~V.}\ \bibnamefont
  {Zhukov}}, \bibinfo {author} {\bibfnamefont {B.~V.}\ \bibnamefont {Danilin}},
  \bibinfo {author} {\bibfnamefont {D.~V.}\ \bibnamefont {Fedorov}}, \bibinfo
  {author} {\bibfnamefont {J.~M.}\ \bibnamefont {Bang}}, \bibinfo {author}
  {\bibfnamefont {I.~J.}\ \bibnamefont {Thompson}}, \ and\ \bibinfo {author}
  {\bibfnamefont {J.~S.}\ \bibnamefont {Vaagen}},\ }\href {\doibase
  10.1016/0370-1573(93)90141-Y} {\bibfield  {journal} {\bibinfo  {journal}
  {Phys. Rept.}\ }\textbf {\bibinfo {volume} {231}},\ \bibinfo {pages} {151}
  (\bibinfo {year} {1993})}\BibitemShut {NoStop}%
\bibitem [{\citenamefont {Stoks}\ \emph {et~al.}(1994)\citenamefont {Stoks},
  \citenamefont {Klomp}, \citenamefont {Terheggen},\ and\ \citenamefont
  {de~Swart}}]{Stoks:1994wp}%
  \BibitemOpen
  \bibfield  {author} {\bibinfo {author} {\bibfnamefont {V.~G.~J.}\
  \bibnamefont {Stoks}}, \bibinfo {author} {\bibfnamefont {R.~A.~M.}\
  \bibnamefont {Klomp}}, \bibinfo {author} {\bibfnamefont {C.~P.~F.}\
  \bibnamefont {Terheggen}}, \ and\ \bibinfo {author} {\bibfnamefont {J.~J.}\
  \bibnamefont {de~Swart}},\ }\href {\doibase 10.1103/PhysRevC.49.2950}
  {\bibfield  {journal} {\bibinfo  {journal} {Phys. Rev.}\ }\textbf {\bibinfo
  {volume} {C49}},\ \bibinfo {pages} {2950} (\bibinfo {year} {1994})}
  \BibitemShut {NoStop}%
\bibitem [{\citenamefont {Wiringa}\ \emph {et~al.}(1995)\citenamefont
  {Wiringa}, \citenamefont {Stoks},\ and\ \citenamefont
  {Schiavilla}}]{Wiringa:1994wb}%
  \BibitemOpen
  \bibfield  {author} {\bibinfo {author} {\bibfnamefont {R.~B.}\ \bibnamefont
  {Wiringa}}, \bibinfo {author} {\bibfnamefont {V.~G.~J.}\ \bibnamefont
  {Stoks}}, \ and\ \bibinfo {author} {\bibfnamefont {R.}~\bibnamefont
  {Schiavilla}},\ }\href {\doibase 10.1103/PhysRevC.51.38} {\bibfield
  {journal} {\bibinfo  {journal} {Phys. Rev.}\ }\textbf {\bibinfo {volume}
  {C51}},\ \bibinfo {pages} {38} (\bibinfo {year} {1995})}
  \BibitemShut {NoStop}%
\bibitem [{\citenamefont {Entem}\ and\ \citenamefont
  {Machleidt}(2003)}]{Entem:2003ft}%
  \BibitemOpen
  \bibfield  {author} {\bibinfo {author} {\bibfnamefont {D.~R.}\ \bibnamefont
  {Entem}}\ and\ \bibinfo {author} {\bibfnamefont {R.}~\bibnamefont
  {Machleidt}},\ }\href {\doibase 10.1103/PhysRevC.68.041001} {\bibfield
  {journal} {\bibinfo  {journal} {Phys. Rev.}\ }\textbf {\bibinfo {volume}
  {C68}},\ \bibinfo {pages} {041001} (\bibinfo {year} {2003})}
  \BibitemShut {NoStop}%
\bibitem [{\citenamefont {Gl{\"o}ckle}(1983)}]{Glockle:1983un}%
  \BibitemOpen
  \bibfield  {author} {\bibinfo {author} {\bibfnamefont {W.}~\bibnamefont
  {Gl{\"o}ckle}},\ }\href@noop {} {\emph {\bibinfo {title} {{The quantum
  mechanical few-body problem}}}}\ (\bibinfo  {publisher} {Springer}\ \bibinfo
  {year} {1983})\BibitemShut {NoStop}%
\end{thebibliography}%

\clearpage

\begin{table}[ht]
\centering
\begin{tabular}{cccc}

\hline\hline
&&\\
{ label}  & rank &support energy $E_l$ [MeV] & \hspace{1.6cm}support momenta $k_l$ [fm$^{-1}$]\hspace{1cm}\\
&&\\
 \hline
&&\\
EST3-1&3&$-35,-15,-5$ &$0.92,\;0.60,\;0.35 $ \\
EST3-2&3&$-5,-5,-5,$ &$0.8,\;1.1,\;2.5$  \\
EST3-3&3&$-25,-5,-5$&$0.8,\;0.8,\;1.1 $  \\
EST3-4&3&$-65,-10,-10$&$0.8,\;0.8,\;2.5  $  \\
&&\\
%
EST4-1&4&$-20,-3,-3,-3$ &$0.7,\;0.4,\;1.1,\;2.5 $\\
EST4-2&4&$-40,-3,-3,-3 $ &$0.7,\;0.4,\;1.1,\;2.5$  \\
EST4-3&4&$-40,-5,-5,-5$  &$0.7\;,0.4,\;1.1,\;2.5$ \\
EST4-4&4&$-60,-7,-7,-7 $ &$0.7,\;0.4,\;1.1,\;2.5$\\
&&\\
%
EST5-1&5&$-30,-20,-3,-3,-3$ &$0.4,\;0.4,\;0.4,\;1.5,\;2.5 $  \\
EST5-2&5&$-60,-60,-3,-3,-3$&$0.5,\;0.5,0.4,\;1.5,\;2.5$  \\
EST5-3&5&$-40,-30,-5,-5,-5$&$0.3,\;0.3,0.4,\;1.5,\;2.5  $  \\
EST5-4&5&$-60,-40,-5,-5,-5 $ &$0.3,\;0.3,0.4,\;1.5,\;2.5 $  \\
&&\\
%
EST6-1&6&$-20,-20,-20,-3,-3, -3$ &$0.4,\;1.1,\;2.5,\;0.4,\;1.1,\;2.5 $ \\
EST6-2&6&$-30,-30,-30,-3,-3, -3$ &$0.4,\;1.1,\;2.5,\;0.4,\;1.1,\;2.5 $\\
EST6-3&6&$-40,-40,-40,-5,-5, -5 $ &$0.4,\;1.1,\;2.5,\;0.4,\;1.1,\;2.5 $   \\
EST6-4&6&$-60,-60,-60,-5,-5, -5$ &$0.4,\;1.1,\;3.5,\;0.4,\;1.1,\;2.5 $\\
&&\\
EST7-1&7&$-20,-20,-20,-3,-3,-3, -3 $ &$0.4,\;1.1,\;3.0,\;0.4,\;1.1,\;3.0,\;15.0 $\\
EST7-2&7&$-30,-30,-30,-3,-3,-3, -3$&$0.4,\;1.1,\;3.0,\;0.4,\;1.1,\;2.5,\;15.0 $\\
EST7-3&7&$-40,-40,-40,-40,-5,-5, -5 $&$0.4,\;1.1,\;3.0,\;15.0,\;0.4,\;1.1,\;3.0$\\
EST7-4&7&$-60,-60,-60,-60,-5,-5, -5$ &$0.4,\;1.1,\;3.0,\;15.0,\;0.4,\;1.1,\;3.0$\\
&\\
EST8-1&8&$-60,-60,-60,-60,-5,-5,-5, -5$ &$0.4,\;1.1,\;3.0,\;15.0,\;0.4,\;1.1,\;3.0,\;15.0$\\
&\\
\hline

\end{tabular}
\caption{Separable representations of the CD-Bonn potential~\cite{Machleidt:2000ge}  in the
energy range $-\infty$~$<$~$E_{2b}$~$\le-2$~MeV. The  labels and ranks of the
separable potentials are listed in the first and second column. The
corresponding support energies and momenta  are given in the third
and fourth columns.    
  }
\label{table:est1}
\end{table}

\begin{table}[ht]
\centering
\begin{tabular}{cccc}

\hline\hline
&&\\
{ label}  & rank &support energy $E_l$ [MeV] & support momenta $k_l$ [fm$^{-1}$]\\
&&\\
 \hline
&&\\
EST3-1&3&$-20,-1,-1$ &$0.5\;0.8,\;1.2 $ \\
EST3-2&3&$-30,-2,-2,$ &$0.5,\;0.8,\;1.2$  \\
EST3-3&3&$-15,-3,-3$&$0.4,\;0.4,\;2.0 $  \\
EST3-4&3&$-5,-5,-5$&$0.5,\;1.2,\;2.0  $  \\
&&\\
%
EST4-1&4&$-20,-5,-5,-5$ &$0.3,\;0.8,\;1.2,\;2.0 $\\
EST4-2&4&$-30,-8,-8,-8 $ &$0.4,\;0.8,\;1.2,\;2.0$  \\
EST4-3&4&$-40,-12,-12,-12$  &$0.4,\;0.8,\;1.2,\;2.0$ \\
EST4-4&4&$-40,-15,-15,-15 $ &$0.4,\;0.8,\;1.2,\;2.0 $\\
&&\\
%
EST5-1&5&$-40,-20,-5,-5,-5$ &$0.3,\;0.5,\;0.8,\;1.2,\;2.0  $  \\
EST5-2&5&$-60,-30,-10,-10,-10$&$0.3,\;0.5,\;0.8,\;1.2,\;2.0 $  \\
EST5-3&5&$-40,-40,-3,-3,-3$&$0.4,\;1.0,\;0.8,\;1.2,\;2.0 $  \\
EST5-4&5&$-60,-40,-20,-20,-20 $ &$0.4,\;0.4,\;0.8,\;1.2,\;2.0 $  \\
&&\\
%
EST6-1&6&$-20,-20,-20,-5,-5, -5$ &$0.8,\;1.1,\;2.0,\;0.8,\;1.2,\;2.0 $ \\
EST6-2&6&$-20,-20,-20,-8,-8, -8$ &$\;0.8,\;1.1,\;2.0,\;0.4,\;1.1,\;2.0 $\\
EST6-3&6&$-30,-30,-30,-10,-10, -10 $ &$\;0.8,\;1.1,\;3.0,\;0.4,\;1.1,\;2.0 $   \\

EST6-4&6&$-40,-40,-40,-15,-15, -15$ &$\;0.8,\;1.1,\;3.0,\;0.4,\;1.1,\;2.0 $\\
&\\
EST7-1&7&$-30,-30,-30,-30,-5,-5, -5$ &$\;0.8,\;1.1,\;2.0,\;3.2,\;0.4,\;1.1,\;2.0$\\
EST7-2&7&$-40,-40,-40,-40,-5,-5, -5$ &$\;0.8,\;1.1,\;2.0,\;3.2,\;0.4,\;1.1,\;2.0$\\
EST7-3&7&$-50,-50,-50,-50,-10,-10, -10 $ &$\;0.8,\;1.1,\;2.0,\;3.2,\;0.4,\;1.1,\;2.2$\\
EST7-4&7&$-60,-60,-60,-60,-10,-10, -10$ &$\;0.8,\;1.1,\;2.0,\;3.2,\;0.4,\;1.1,\;3.2$\\
&\\
%
EST8-1&8&$-20,-20,-20,-20,-5,-5, -5,-5$ & $\;0.8,\;1.1,\;2.0,\;3.2,\;0.4,\;1.1,\;2.0,\;3.2 $\\
EST8-2&8&$-30,-30,-30,-30,-8,-8, -8,-8$ &$\;0.8,\;1.1,\;2.0,\;3.2,\;0.4,\;1.1,\;2.0,\;3.2 $\\
EST8-3&8&$-50,-50,-50,-50,-10,-10, -10,-10 $ &$\;0.8,\;1.1,\;2.0,\;3.2,\;0.4,\;1.1,\;2.0,\;3.2 $\\
EST8-4&8&$-60,-60,-60,-60,-10,-10, -10,-10$ &$\;0.8,\;1.1,\;2.0,\;3.2,\;0.4,\;1.1,\;2.0,\;3.2 $\\
&\\
\hline

\end{tabular}
\caption{Separable representations of the Bang potential~\cite{Bang:1979ihm}  in the
energy range $-\infty$~$<$~$E_{2b}$~$\le-2$~MeV. The  labels and ranks of the
separable potentials are listed in the first and second column. The
corresponding support energies and momenta  are given in the third
and fourth columns.      
  }
      \label{table:est2}
\end{table}

\begin{table}[ht]
\centering
\begin{tabular}{cccc}

\hline\hline
&&\\
label& rank&$E_{3b}$ [MeV]\\
&&\\ 
\hline
&&\\
EST3-1& 3 &-3.7967\\
EST3-2& 3 &-3.7519\\ 
EST3-3& 3 &-3.7507\\ 
EST3-4& 3 &-3.7480\\   
&\\
EST4-1& 4 &-3.7774\\
EST4-2& 4 &-3.7737\\ 
EST4-3& 4 &-3.7712\\ 
EST4-4& 4 &-3.7823\\   
&&\\
EST5-1& 5 &-3.7847\\
EST5-2& 5 &-3.7848\\ 
EST5-3& 5 &-3.7855\\ 
EST5-4& 5 &-3.7845\\   
&&\\
EST6-1& 6 &-3.7867\\
EST6-2& 6 &-3.7868\\ 
EST6-3& 6 &-3.7871\\ 
EST6-4& 6 &-3.7870\\   
&&\\
EST7-1& 7 &-3.7867\\
EST7-2& 7 &-3.7867\\ 
EST7-3& 7 &-3.7867\\ 
EST7-4& 7 &-3.7867\\   
&\\
\hline
EXACT&  &-3.787\\
\hline

\end{tabular}
\caption{Three-body binding energies for the ground state of $^{6}$Li calculated using the separable
representations of the CD-Bonn potential listed in Table~\ref{table:est1}. 
The labels and ranks of the separable potentials are 
shown in the first and second column. The corresponding three-body binding energies are listed in the third column. The EST8-4 separable representation of the Bang potential defined in Table~\ref{table:est2} 
is adopted for the $n\alpha$
subsystem.  Calculations shown in this table do not include 
the Coulomb potential.   
  }
      \label{table:eb1}
\end{table}

\begin{table}[ht]
\centering
\begin{tabular}{cccc}

\hline\hline
&&\\
label & rank&$E_{3b}$ [MeV]\\
&&\\ 
\hline
&&\\
EST3-1& 3 &-3.7527\\
EST3-2& 3 &-3.7524\\ 
EST3-3& 3 &-3.7151\\ 
EST3-4& 3 &-3.7127\\   
&\\
EST4-1& 4 &-3.7788\\
EST4-2& 4 &-3.7777\\ 
EST4-3& 4 &-3.7773\\ 
EST4-4& 4 &-3.7778\\   
&&\\
EST5-1& 5 &-3.7798\\
EST5-2& 5 &-3.7797\\ 
EST5-3& 5 &-3.7807\\ 
EST5-4& 5 &-3.7806\\    
&&\\
EST6-1& 6 &-3.7856\\
EST6-2& 6 &-3.7852\\ 
EST6-3& 6 &-3.7852\\ 
EST6-4& 6 &-3.7856\\   
&&\\
EST7-1& 7 &-3.7868\\
EST7-2& 7 &-3.7864\\ 
EST7-3& 7 &-3.7867\\ 
EST7-4& 7 &-3.7865\\   
&\\
&&\\
EST8-1& 8 &-3.7870\\
EST8-2& 8 &-3.7870\\ 
EST8-3& 8 &-3.7866\\ 
EST8-4& 8 &-3.7868\\   
&\\
\hline
EXACT&  &-3.787\\
\hline

\end{tabular}
\caption{Three-body binding energies for the  ground state of $^{6}$Li calculated using the separable representations of the Bang potential listed in Table~\ref{table:est2}. The labels and ranks of the separable potentials are 
shown in the first and second columns. The corresponding three-body binding energies are listed in the third column. The EST8-1 separable representation of the CD-Bonn  potential defined in Table~\ref{table:est1} is adopted for the $np$
subsystem. Calculations shown in this table do not include 
the Coulomb potential.
  }
      \label{table:eb2}
\end{table}

\begin{table}
\begin{tabular}{|r|c|c|c| }
\hline\hline
  $\lambda$ [fm$^{-1}$]   & $E (\lambda)$ [MeV]  & $\langle E (\lambda) \rangle$ [MeV]  & 
$ {\mathcal P}_{TDB}(\lambda) $ [$\%$]\\ 
\hline \hline
 0       & -35.65  & -35.65  &    89.21       \\ \hline
 0.01    & -32.15  & -35.62  &    88.08         \\ \hline
 0.1     & -4.798  & -16.84  &    30.52           \\ \hline
 1       & -3.842  & -3.886  &    1.133$\times 10^{-2}$     \\ \hline
 10      & -3.794  & -3.801  &    1.654$\times 10^{-4}$          \\ \hline
 100     & -3.788  & -3.789  &    1.765$\times 10^{-6}$          \\ \hline
 1000    & -3.787  & -3.788  &    1.843$\times 10^{-8}$             \\ \hline
 10000   & -3.787  & -3.787  &    2.450$\times 10^{-10}$          \\ \hline
 100000  & -3.787  & -3.787  &    2.328$\times 10^{-11}$             \\ \hline
$\infty$ & -3.787  & -3.787  &    1.259$\times 10^{-10}$          \\ \hline \hline
\end{tabular}
\caption{The binding energy of the ground state of $^6$Li computed with different 
values of the parameter $\lambda$ in the projection operator. For this calculation the 
Coulomb interaction in the $p\alpha$ subsystem is omitted. 
The quantity $\langle E (\lambda) \rangle$ represents the expectation value of Hamiltonian
computed according to Eq.~(\ref{eq:expect}) with the corresponding projection. 
The probability 
${\mathcal P}_{TDB}(\lambda)$ defined in Eq.~(\ref{eq:PFstate}) for finding the Pauli
forbidden S$_{1/2}$ state in the $^6$Li ground state wave function is given in the
last column.
}
\label{table-lam}
\end{table}

\clearpage

\newpage

\noindent
\begin{figure}
\begin{center}
\includegraphics[scale=.55]{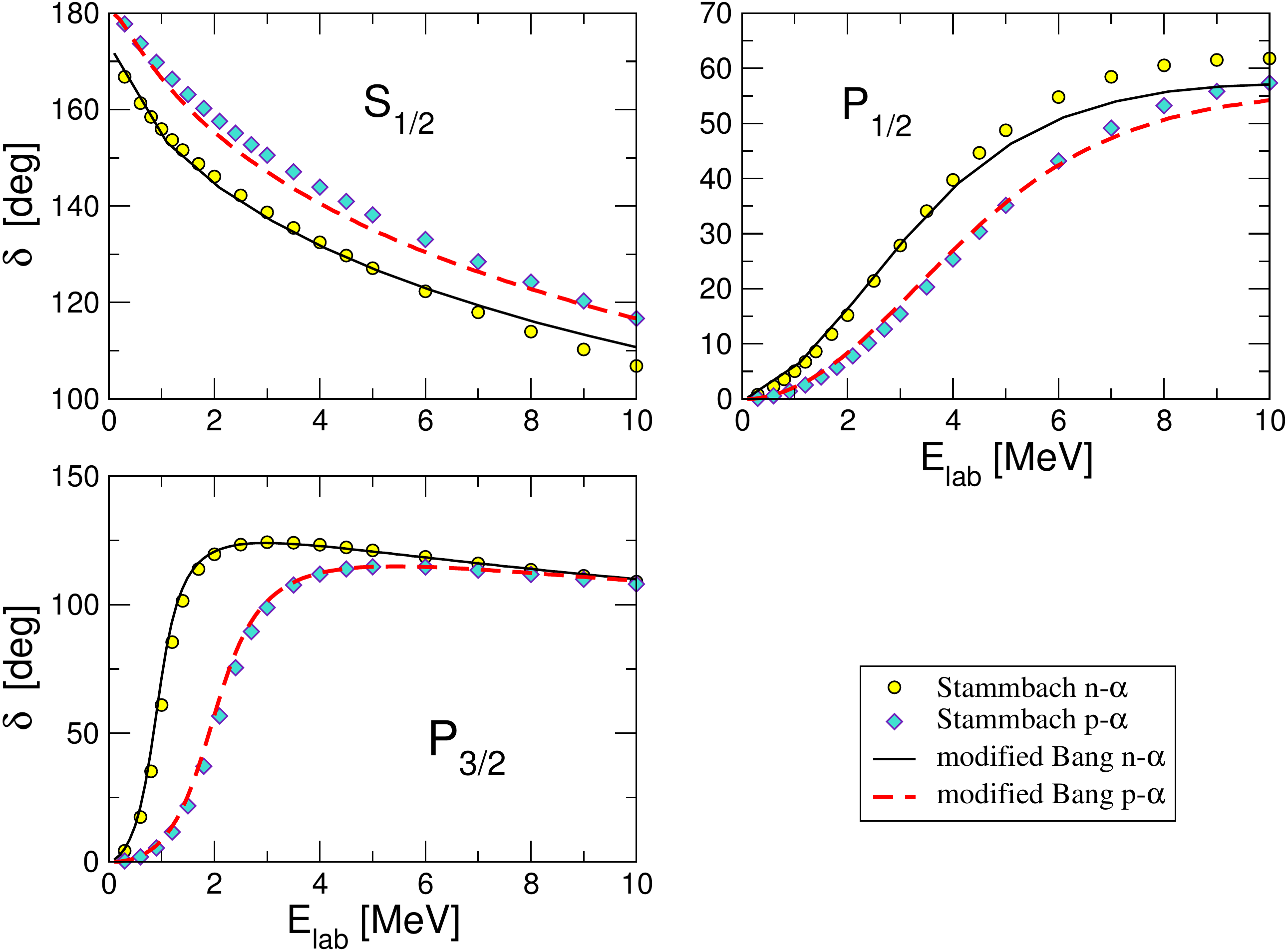}
\vspace{3mm}
\caption{The S- and P-wave phase shifts in the $n\alpha$ and $p\alpha$ subsystems as function of the 
neutron/proton laboratory kinetic energy. The solid (dashed) lines represent the calculations
with the modified Bang interaction~\cite{Bang:1983xpz} for the
$n\alpha$ and $p\alpha$ systems. The phase-shifts extracted from an
R-matrix fit~\cite{Stammbach72} are shown for $n\alpha$ by filled circles and for $p\alpha$ by
filled diamonds.
}
\label{fig1}
\end{center}
\end{figure}

\begin{figure}[ht]
\centering
\includegraphics[width=15cm]{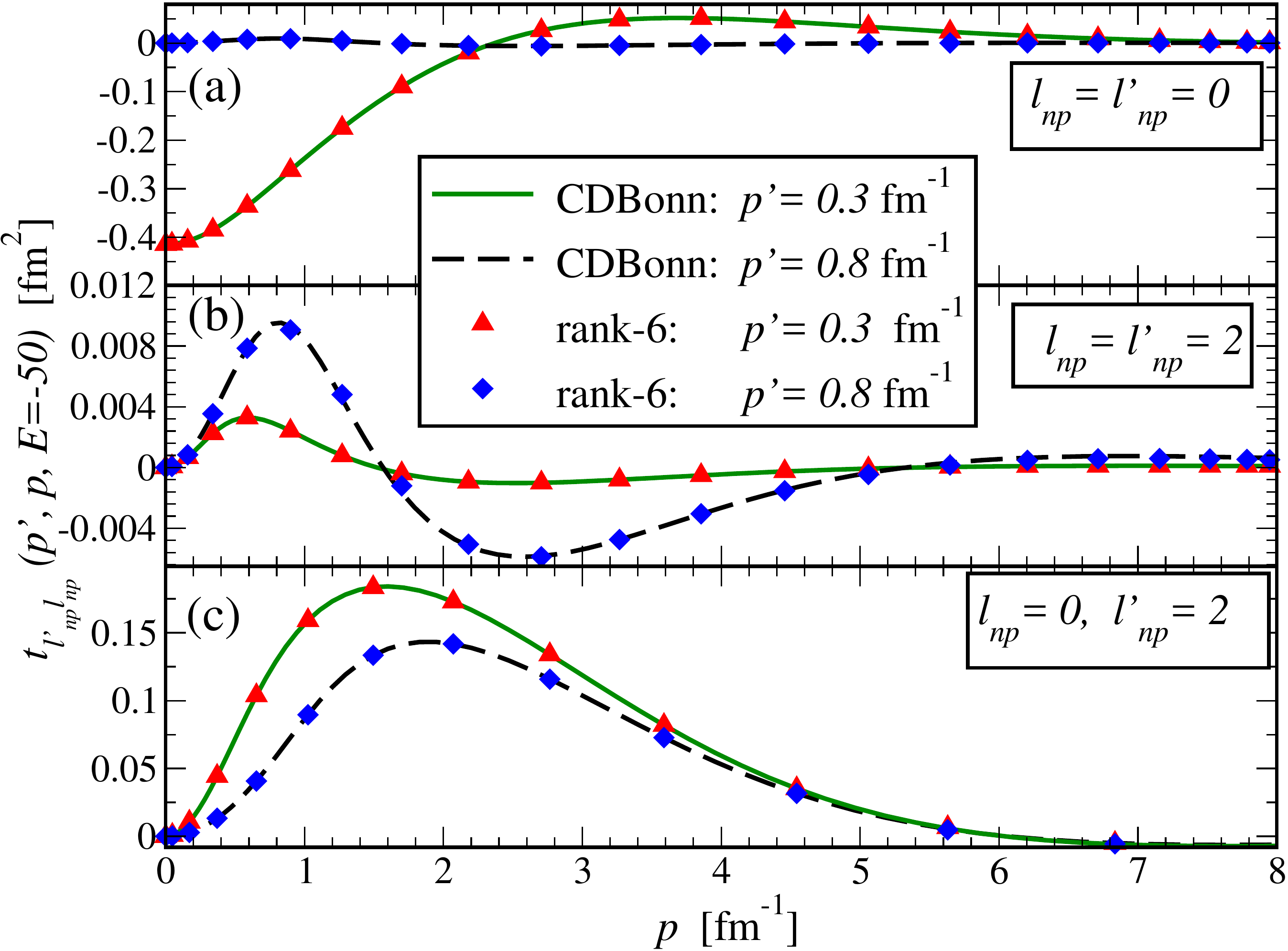}
\caption{The  off-shell $t$~matrix elements $t_{l_{np}'l_{np}}~(p',p;E)$ for the $np$ system as function of the off-shell momentum $p$. 
The center of mass energy is $E_{2b}$=~-50~MeV while the
total angular momentum and spin are fixed at $J_{np}=S_{np}=1$.
The $t$~matrix elements $t_{00}~(p',p;E)$ are shown in panel (a) while the ones corresponding to $l_p=l_p'=2$ are illustrated in panel
(b). Panel (c) shows the matrix elements $t_{20}~(p',p;E)$.   The solid
and dashed lines depict the $t$~matrix elements computed with the CD-Bonn potential for $p'$= 0.3~fm$^{-1}$ and $p'$=~0.8~fm$^{-1}$
respectively. 
The results obtained using a rank-6 separable representation of the CD-Bonn potential are indicated by upward triangles for $p'$=~0.3~fm$^{-1}$ and
by diamonds for $p'$=~0.8~fm$^{-1}$.
The six EST support points are located at
$\{E_l,p_l\}~=\{-60,0.4\},\{-60,1.1\},\{-60,2.5\},\{-5,0.4\} ,\{-5,1.1\},\{-5,2.5\}$. The energies have
units of MeV while the momenta are given in fm$^{-1}$.        
}
\label{fig:3.2.1}
\end{figure}

\noindent
\begin{figure}
\begin{center}
\includegraphics[scale=.55]{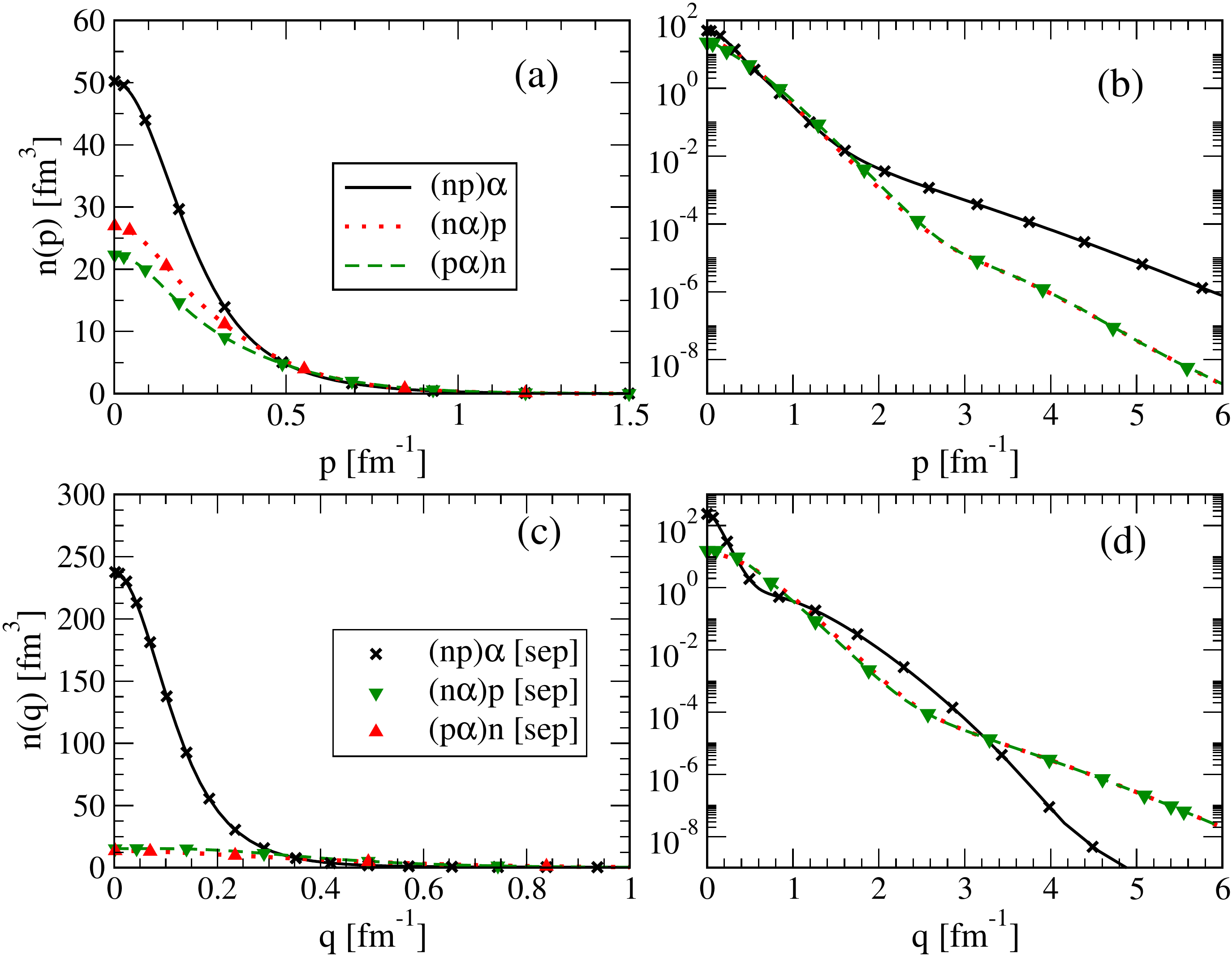}
\vspace{3mm}
\caption{Panels (a) and (b) show the momentum distributions $n(p)$ in the $(np)$-$\alpha$ (solid
line), $(n\alpha)$-$p$ (dotted line), and $(p\alpha)$-$n$ (dashed line) arrangement channels of the
$^6$Li ground state calculated with the CD-Bonn~\cite{Machleidt:2000ge} $np$ interaction and the
modified Bang~\cite{Bang:1979ihm} $n\alpha$ interaction. For the $p\alpha$ interaction the
Coulomb interaction given in Eq.~(\ref{eq:Coulomb}) is added. Panels (c) and (d) show the momentum 
distributions $n(q)$ for the same arrangement channels as panels (a) and (b). The crosses as well as the upward
and downward triangles correspond to the same calculations but using separable forces.
}
\label{fig2}
\end{center}
\end{figure}

\noindent
\begin{figure}
\begin{center}
\includegraphics[scale=.55]{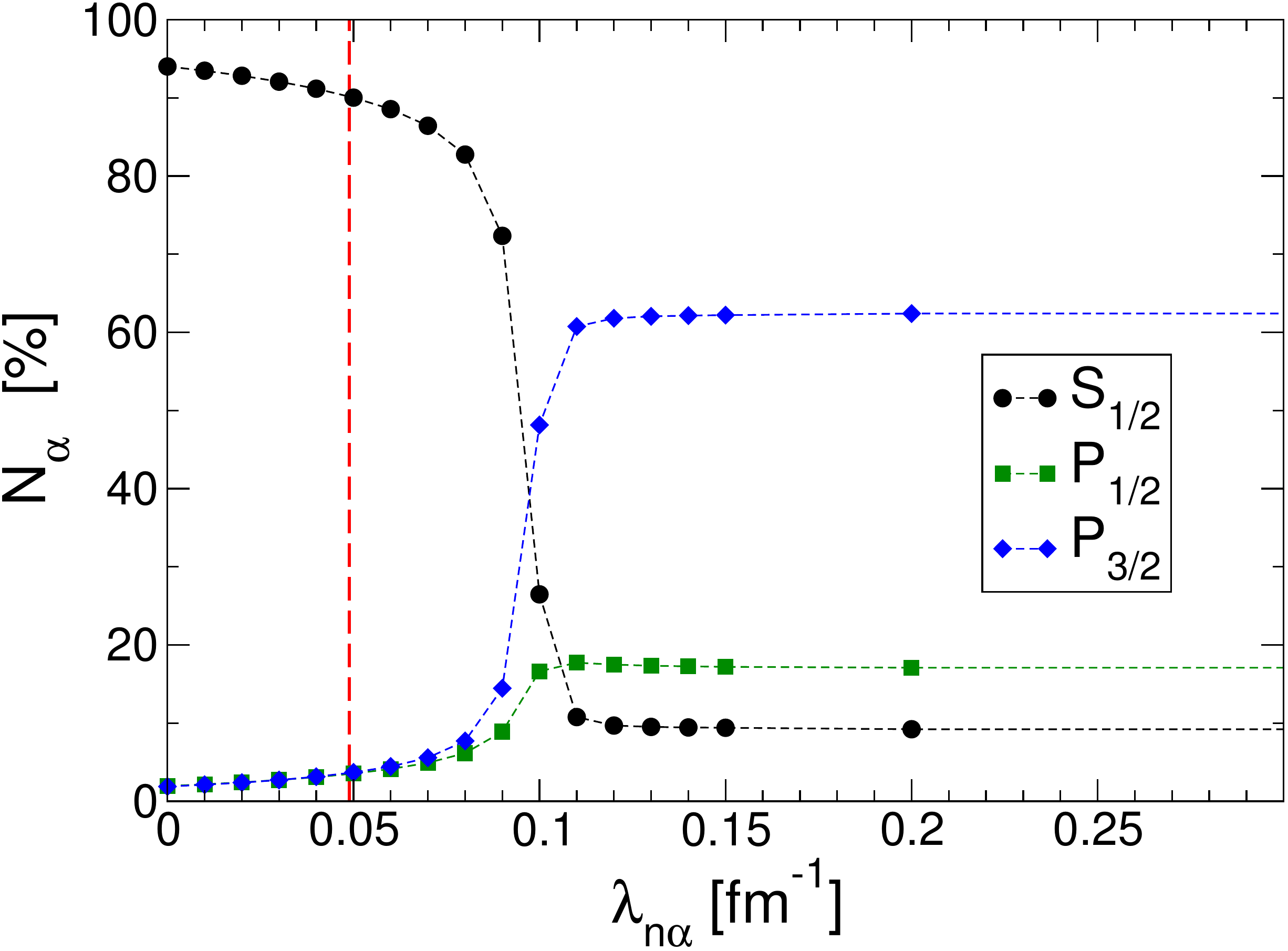}
\vspace{3mm}
\caption{The probability $N_{\beta}(\lambda)$ for the $^6$Li three-body ground state as a function of the projector strength $\lambda$ calculated according Eq.~(\ref{eq:probability}). The solid, dashed, and dot-dashed lines represent the $S_{1/2}$, $P_{1/2}$ and $P_{3/2}$ partial wave states of the $n\alpha$ subsystem.
The dashed vertical line indicates the value of $\lambda$ at which the $S_{1/2}$ state becomes unbound.
}
\label{fig3}
\end{center}
\end{figure}

\end{document}